\documentclass[5p, times,authoryear]{tp}
\usepackage{amssymb}
\usepackage{latexsym}
\usepackage{amsmath}
\usepackage[table]{xcolor}

\usepackage{url}
\usepackage{xcolor}
\usepackage{hyperref}

\definecolor{newcolor}{rgb}{.8,.349,.1}

\begin{document}

\begin{frontmatter}

\title{Shape constrained CNN for segmentation guided prediction of myocardial shape and pose parameters in cardiac MRI}

\author[1,3]{Sofie Tilborghs\corref{cor1}}
\cortext[cor1]{Corresponding author at: KU Leuven, 3000 Leuven, Belgium.}
\ead{sofie.tilborghs@kuleuven.be}
\author[2,3]{Jan Bogaert}
\author[1,3]{Frederik Maes}

\address[1]{Department of Electrical Engineering, ESAT/PSI, KU Leuven, Leuven, Belgium}
\address[2]{Department of Imaging and Pathology, Radiology, KU Leuven, Leuven, Belgium}
\address[3]{Medical Imaging Research Center, UZ Leuven, Herestraat 49 - 7003, Leuven, 3000, Belgium}

\begin{abstract}
%%%
Semantic segmentation using convolutional neural networks (CNNs) is the state-of-the-art for many medical image segmentation tasks including myocardial segmentation in cardiac MR images. However, the predicted segmentation maps obtained from such standard CNN do not allow direct quantification of regional shape properties such as regional wall thickness. Furthermore, the CNNs lack explicit shape constraints, occasionally resulting in unrealistic segmentations. In this paper, we use a CNN to predict shape parameters of an underlying statistical shape model of the myocardium learned from a training set of images. Additionally, the cardiac pose is predicted, which allows to reconstruct the myocardial contours. The integrated shape model regularizes the predicted contours and guarantees realistic shapes. We enforce robustness of shape and pose prediction by simultaneously performing pixel-wise semantic segmentation during training and define two loss functions to impose consistency between the two predicted representations: one distance-based loss and one overlap-based loss. We evaluated the proposed method in a 5-fold cross validation on an in-house clinical dataset with 75 subjects and on the ACDC and LVQuan19 public datasets. We show the benefits of simultaneous semantic segmentation and the two newly defined loss functions for the prediction of shape parameters. Our method achieved a correlation of 99$\%$ for left ventricular (LV) area on the three datasets, between 91$\%$ and 97$\%$ for myocardial area, 98-99$\%$ for LV dimensions and between 80$\%$ and 92$\%$ for regional wall thickness.
%%%%
\end{abstract}


\end{frontmatter}

%\linenumbers

\section{Introduction}
Segmentation of cardiac structures such as the myocardium in magnetic resonance (MR) images is a prior step for quantitative assessment of cardiac structure and function. Relevant measures include left ventricular (LV) volume, ejection fraction (EF) and myocardial wall thickness. This segmentation is performed mostly manually in clinical practice, which is time-consuming and subject to inter- and intra-observer variation, explaining the large amount of research for automated methods. During the last years, convolutional neural networks (CNNs) have shown to outperform traditional model-based segmentation techniques and quickly became the method of choice for this task (\cite{Bernard2018}). However, the predicted segmentation maps obtained from a standard CNN only allow automatic quantification of global shape properties and not of regional shape properties such as regional wall thickness which would additionally require the pose of the heart. Furthermore, since conventional CNNs are trained to predict a class probability for each voxel, they are missing explicit shape constraints, occasionally resulting in unrealistic segmentations with missing or disconnected regions (\cite{Khened2019,Painchaud2019}). Therefore, several authors have proposed to integrate a shape prior in their CNN in the form of (1) atlases (\cite{Duan2018a,Zotti2019,Lee2019}), (2) statistical shape models (\cite{Attar2019,Bhalodia2018,Adams2020,Milletari2017,Schock2020,Wang2018}), (3) hidden representations (\cite{Oktay2018,Yue2019,Painchaud2019,Chen2020a,Biffi2020}) or (4) distance-based shape representations (\cite{Karimi2020,Kervadec2018,Xue2019,Navarro,Dangi2019,Ma2020,Li2020,Zhang2020a,Hu2020}).

The first approach to integrate a shape prior in a CNN is to use an atlas or template image as additional input for the CNN. In \cite{Duan2018a}, a CNN is used to simultaneously predict segmentation and relevant landmarks that are subsequently affinely aligned with landmarks of a set of atlas images to select the most corresponding atlases. The selected atlases are then non-rigidly aligned to the predicted segmentation and labels of the warped atlases are fused to obtain the final segmentation. \cite{Zotti2019} constructed a fuzzy shape prior representing the probability for a voxel to be part of the object of interest. This shape prior is integrated in a segmentation CNN which also predicts the object's center position. \cite{Lee2019} obtain a shape-regularized segmentation by transforming a shape prior with a predicted deformation field. Additionally, they showed the benefit of adding the shape prior as second input in a standard U-Net architecture.

The second approach combines earlier successful methods such as active shape models (ASM, \cite{Cootes1995}) with CNNs. In contrast to CNNs, an ASM constructs a landmark-based statistical shape model from a training dataset by performing principal component analysis (PCA) and fits this model to a new image using learned local intensity models for each landmark, yielding patient-specific global shape coefficients. Regression of these shape coefficients using a CNN is performed by \cite{Bhalodia2018}. \cite{Adams2020} extended this work to a probabilistic approach to quantify uncertainty. Probabilistic surface prediction with a PCA shape prior was also performed by \cite{Tothova2020}. \cite{Attar2019} used both CMR images and patient metadata to predict shape coefficients. In \cite{Milletari2017}, a multi-resolution approach is proposed where the position of coarsely predicted landmarks is refined using a shallow CNN on high resolution patches around the current landmarks. \cite{Schock2020} also use shape coefficient regression in a multi-step approach for knee segmentation. These regression CNNs can be trained by either imposing a loss on the shape coefficients directly (\cite{Attar2019,Bhalodia2018}) or by a loss on the reconstructed landmarks (\cite{Milletari2017,Schock2020}). \cite{Wang2018} create a statistical shape model by performing PCA on signed distance maps. A case-specific model instance is generated by solving a level-set function using an initial segmentation as image force. This model instance is an additional input for a second segmentation CNN. Combining a traditional segmentation approach with deep learning was also performed by \cite{Zhang2020}. In their method, they combine a CNN with an active contour model (\cite{Kass1988}): a CNN is used for initial segmentation prediction and prediction of two parameter maps required for the active contour model. Subsequently, the active contour model refines the segmentation.

The third approach is based on hidden representations of shape. \cite{Oktay2018} use the latent space of an auto-encoder (AE) as shape representation and regularize the shape of predicted segmentations by imposing a similarity loss between AE encodings of ground truth and predicted segmentations. \cite{Yue2019} add an AE, pre-trained on realistic segmentations, after a segmentation network. The dissimilarity between predicted segmentation and AE output is used as additional loss. \cite{Chen2019} calculate shape priors in a multi-view auto-encoder and concatenate the learned latent shape representations to the feature maps of a segmentation CNN. In \cite{Painchaud2019}, a variational auto-encoder (VAE) is used to generate a latent shape space. After initial segmentation, the erroneous segmentation is fed to the VAE encoder to generate a shape vector. The closest point of the latent shape space to this vector is used as corrected shape and is fed to the VAE decoder to obtain the final segmentation. \cite{Chen2020a} also incorporate a shape prior by encoding shape via a VAE. \cite{Biffi2020} employ the ladder VAE framework (\cite{Sonderby2016}), generating latent spaces at different resolutions.

In a fourth approach, implicit regularization of shape is performed through the inclusion of distance-based loss functions in conjunction with traditional overlap-based metrics. \cite{Karimi2020} proposed three new loss functions derived from Hausdorff distance (HD), making a different trade-off between accuracy of HD estimation and implementation efficiency. Their most accurate but most time-consuming method uses unsigned distance maps calculated from ground truth and predicted segmentation maps. Distance maps are also used for many other distance-based loss functions. The boundary loss of \cite{Kervadec2018} and the spatial encoding loss of \cite{Li2020} only require ground truth distance maps and can thus be calculated more efficiently. Other methods (\cite{Xue2019,Tilborghs2020b}) directly predict a signed distance map and use a smooth approximation to the Heaviside function to obtain a segmentation map suitable for conventional overlap-based losses. While previous methods assure full correspondence between segmentation and distance maps, this is not the case for methods that predict both representations using separate decoders (\cite{Dangi2019}) or separate heads (\cite{Navarro}). \cite{Navarro} additionally predict a segmentation map for object edges. \cite{Li2020a} enlarge a conventional segmentation CNN by introducing few convolutional layers after the probabilistic output to obtain a distance map. \cite{Ma2020} compare several of the aforementioned distance losses. They conclude that whereas the inclusion of distance-based loss functions effectively reduces HD, the optimal definition of such loss function is not clear. The use of distance maps and associated boundary losses is more complicated when a patch-based approach is used. In \cite{Zhang2020a}, a 3D+2D approach is proposed to use the boundary loss of \cite{Kervadec2018} for segmentation in large 3D volumes. Instead of distance maps, \cite{Hu2020} jointly predict a segmentation map and a boundary map using two decoders with skip connections. An additional loss imposes consistency between the boundary map and boundaries generated from the predicted segmentation map.

Other approaches attempting to regularize shape are the methods of \cite{Ye2020a}, who modify the lowest resolution of a U-Net architecture to jointly predict a point cloud, and \cite{Liu2020} who introduce a new shape constrained loss function based on the shapes' Hu moments (\cite{Flusser2000}).

In this paper, we propose a hybrid approach for myocardial segmentation in cardiac MR where elements of approaches 2, 3 and 4 are combined to directly obtain a shape representation suitable for regional shape quantification. Similar to approach 2, we perform regression of shape coefficients of a landmark-based statistical shape model, which allow straightforward calculation of regional shape properties. Compared to other methods, we enforce robustness of shape prediction by simultaneously performing semantic segmentation. Similar to approach 3, this is implemented as an encoder-decoder CNN where the most downsampled feature maps are expected to contain relevant shape information. These feature maps are jointly used for separate regression and semantic segmentation branches. Combining segmentation with regression in a similar way in multi-task learning problems was used by \cite{Vigneault2018}, \cite{Zotti2019} and \cite{Yue2019} to estimate cardiac pose, by \cite{Gessert2020} and \cite{Tilborghs2020a} to perform direct quantification of LV parameters and by \cite{Cao2018} for simultaneous hippocampus segmentation and clinical score regression from brain MR images. In our approach, we explicitly enforce consistency between predicted shape and predicted semantic segmentation by introducing two new loss functions, one minimizing the distance between the contours of the two representations and the other maximizing their overlap. Furthermore, we implemented the semantic segmentation task as the prediction of signed distance maps (approach 4).

Since we address the segmentation problem in this paper as a regression problem predicting both shape and pose parameters in order to automatically extract global and regional shape properties, this work is closely related to literature focusing on LV parameter estimation in cardiac MRI. We distinguish three categories of automated methods for this task: (1) methods that extract the parameters from a prior segmentation (\cite{Acero,Khened2019,Zheng2019}), (2) methods that perform direct parameter regression (\cite{Xue2018,Liu2021a}) and (3) methods that perform both segmentation and parameter regression (\cite{Dangi2019a, Yan2019, Xue2017, Gessert2020};\linebreak \cite{Tilborghs2020a}). \cite{Khened2019} extract global cardiac parameters from a segmentation of LV cavity, \linebreak myocardium and right ventricle (RV) automatically generated with a CNN. \cite{Zheng2019} also perform a multi-class segmentation and use the centroids of predicted LV and RV to define the cardiac orientation in order to additionally calculate regional LV parameters. The former approach to define cardiac orientation could not be used in the work of \cite{Acero} since their training dataset did not contain ground truth segmentations of RV. Hence, they pose-normalized the images by registering each image to an atlas. The pose-normalized segmentations then allowed to extract both global and regional LV parameters. Direct regression of multiple LV parameters simultaneously using custom CNNs has been performed by \cite{Xue2018} and \cite{Liu2021a}. However, several studies have shown that LV parameter regression benefits from a joint segmentation (\cite{Xu2019,Tilborghs2020a}). In this respect, \cite{Dangi2019a} and \cite{Yan2019} have added an additional decoder head to a segmentation CNN to perform parameter regression. \cite{Xue2017} proposed a two-step approach where the latent space of a pre-trained auto-encoder is used as input to a shallow CNN for LV parameter regression. Similarly, \cite{Gessert2020} and \cite{Tilborghs2020a} hypothesize that the most downsampled features in an encoder-decoder segmentation CNN provide all relevant information to perform accurate LV parameter estimation. Hence, both approaches added a regression branch at this position. Compared to \cite{Xue2017}, \cite{Gessert2020} and \cite{Tilborghs2020a} simultaneously optimized segmentation and regression branch as a multi-task problem.
  
This paper is an extension of our previous work published in \cite{Tilborghs2020b}. The main novelty in this paper compared to this previous work is the introduction of two losses to enforce consistency between predicted shape and predicted semantic segmentation ($L_{C_c}$ and $L_{C_o}$). We also used an additional landmark loss $L_{p}$ for shape and pose prediction. Furthermore, we provide insight in the prediction of individual shape coefficients (Fig. \ref{fig:shape}). Moreover, we validated our final method on an additional public dataset (ACDC dataset, \cite{Bernard2018}, Tables \ref{tab:RegSegPost}-\ref{tab:physical} and Fig.~\ref{fig:segresult_consistency}) and compare our approach to a state-of-the-art segmentation method (nnUNet, \cite{Isensee2021}, Tables~\ref{tab:litCompGeo}-\ref{tab:litCompPhysical} and Fig.~\ref{fig:segresult_nnunet}).

% === II. METHODS ========================
% =================================================================================
\section{Methods}
The proposed CNN jointly performs regression of shape and pose parameters of an underlying statistical model and semantic segmentation by prediction of signed distance maps. We focus on consistent prediction of both myocardial representations in order to straightforwardly calculate reliable regional shape properties from predicted shape parameters. This section starts with the construction of the shape model (Section \ref{sec:shapemodel}), followed by the CNN architecture and used loss functions in Section \ref{sec:cnn}. Subsequently, the datasets used in this work are described (Section \ref{sec:data}) and implementation (Section \ref{sec:details}) and training (Section \ref{sec:training}) details are provided.  
\subsection{Shape model}
\label{sec:shapemodel}
The myocardium in a short-axis (SA) cross-section is approximated by a set of $N$ endo- and epicardial landmarks radially sampled over uniform angular offsets relative to the LV center position $\textbf{c}$ and anatomical orientation $\theta$ as defined in Section \ref{sec:details}. From a training set of images, a statistical shape model representing the mean shape $\overline{\textbf{s}}$ and the modes of variation is calculated using PCA. A single model is created to represent the shape variation of all individual 2D slices covering the heart from apex to base at multiple time points. For each slice, the myocardial contour $\textbf{p}$, consisting of x- and y-coordinates of the $N$ landmarks, is first normalized, resulting in the pose-normalized shape $\textbf{s}$:
\begin{equation}
\label{eq:pointnorm}
\begin{aligned}
\textbf{s}
&= T_{Pose}(\textbf{p},\textbf{c},\theta)\\
&=
\begin{bmatrix}
\textbf{s}_{x}\\
\textbf{s}_{y}\\
\end{bmatrix}=
\begin{bmatrix}
\cos(\theta) & \sin(\theta)\\
-\sin(\theta) & \cos(\theta)
\end{bmatrix}
\begin{bmatrix}
\textbf{p}_{x}-c_{x}\\
\textbf{p}_{y}-c_{y}
\end{bmatrix}.
\end{aligned}
\end{equation}
The eigenvectors $\textbf{V} = \{\textbf{v}_1,...,\textbf{v}_m,...,\textbf{v}_{2N}\}$ and corresponding eigenvalues $\lambda_m$ are obtained from the singular value decomposition of the covariance matrix of the normalized shapes $\textbf{s}-\overline{\textbf{s}}$. The shape of the myocardium is approximated by the $M$ first eigenmodes with parameters $\textbf{b} = \{b_1,...,b_m\}$:
\begin{equation}
\label{eq:pca}
    \textbf{s} \approx \tilde{\textbf{s}}(\textbf{b}) = \overline{\textbf{s}} + \sum_{m=1}^{M} b_{m} \cdot \sqrt{\lambda_m}\cdot \textbf{v}_m .
\end{equation}
With this formulation, the variance of the distribution of shape coefficients $b_m$ is 1 for every mode $m$. The final myocardial landmarks $\textbf{p}$ used for training the network are obtained by applying the inverse pose-normalization transformation $T_{Pose}^{-1}$ \linebreak (Eq.~\ref{eq:pointnorm}) to the approximated shape $\tilde{\textbf{s}}(\textbf{b})$. The endo- and epicardial contours are obtained by cubic spline interpolation of the landmarks, from which the signed distance map $D$ representing the Euclidean distance to the myocardial boundaries is obtained.  
\subsection{CNN}
\label{sec:cnn}
\begin{figure}[tb] %(figure* voor over de hele breedte van de paper)
	\centering
	\includegraphics[width =\linewidth]{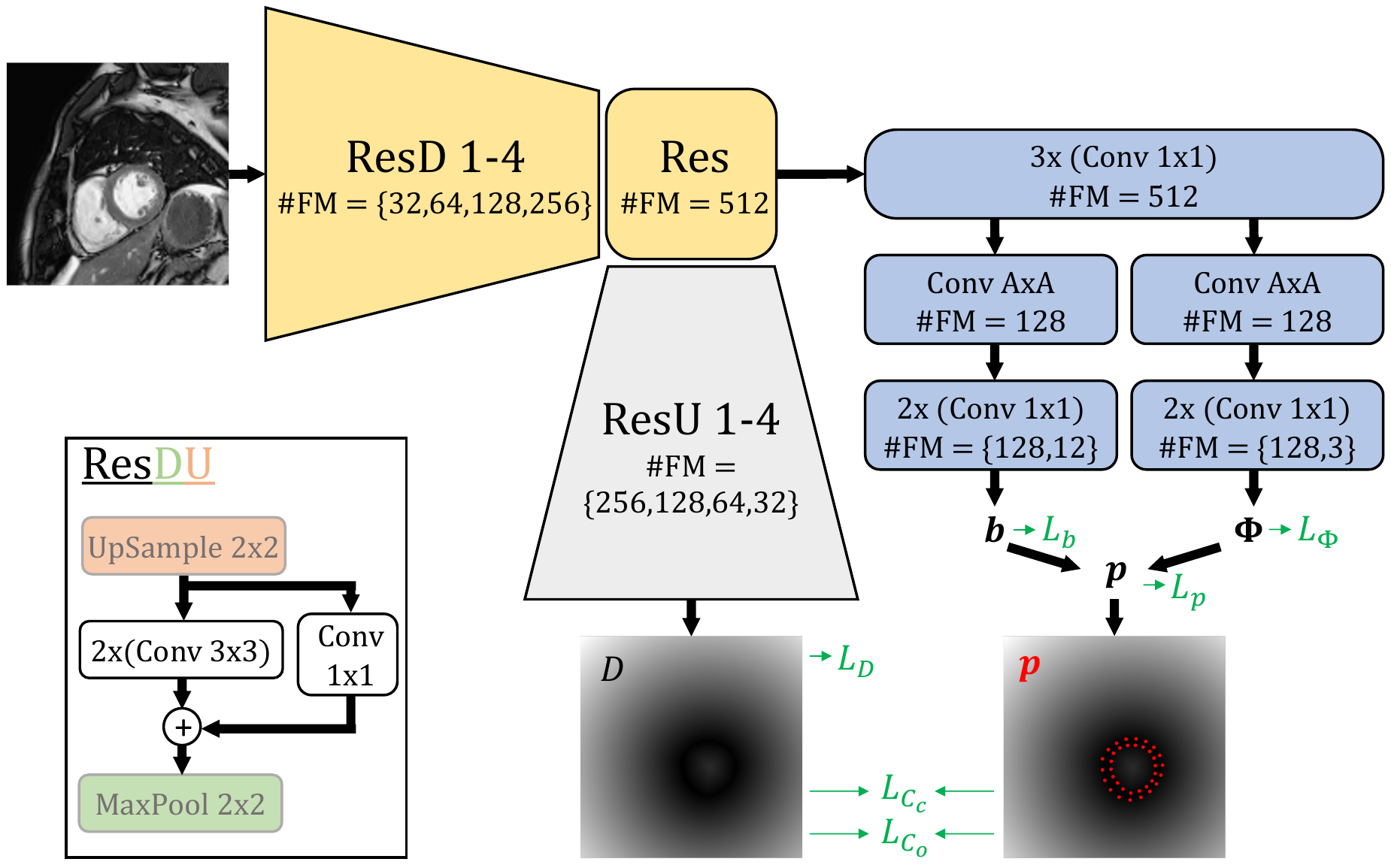}
	\caption{Proposed CNN architecture with three outputs: shape coefficients $\textbf{b}$, pose $\boldsymbol{\Phi}$ and distance map $D$. Landmarks $\textbf{p}$ are calculated from $\textbf{b}$ and $\boldsymbol{\Phi}$. Loss functions are depicted in green. Consistency between $D$ and $\textbf{p}$ is enforced by $L_{C_c}$ and $L_{C_o}$. Details of residual (Res), downsampling Res (ResD) and upsampling Res (ResU) blocks are given on the left. The number of feature maps ($\#FM$) is the same for every Conv layer in one Res block. The filter size $A$ in a Conv layer is equal to the dimensions of that layer's input. } 
	\label{fig:network}
\end{figure}
Fig.~\ref{fig:network} shows the proposed CNN architecture to perform both shape and pose parameter regression and semantic segmentation. The CNN has three outputs: (1) shape coefficients $\textbf{b}$, (2) pose parameters $\boldsymbol{\Phi} = \{\theta,c_{x},c_{y}\}$ and (3) signed distance map $D$. Layers used for parameter regression are depicted in blue, for semantic segmentation in gray and jointly for both tasks in yellow. Semantic segmentation is performed by an encoder-decoder architecture that predicts signed distance maps instead of a conventional binary segmentation (as motivated further). The pathway for parameter regression branches off from the most downsampled, central, feature maps of this\linebreak encoder-decoder. 

The network is trained using a combination of loss functions, either imposing similarity with the ground truth ($L_{b}$, $L_{\Phi}$, $L_{p}$ and $L_{D}$) or imposing consistency between the different outputs ($L_{C_c}$ and $L_{C_o}$). The shape loss $L_{b}$ is defined as the mean squared error (MSE) between ground truth and predicted coefficients $\textbf{b}$:
\begin{equation}
L_{b} = \frac{1}{M} \|\textbf{b}_{t}-\textbf{b}_{p}\|^2,
\end{equation}
where subscripts $t$ and $p$ denote ground truth and prediction, respectively. The pose loss $L_{\Phi}$ is a combination of cosine loss on orientation and MSE on position, weighted with $\mu_{\Phi}$:
\begin{equation}
\begin{aligned}
L_{\Phi} = \left(-\cos(\theta_t-\theta_p)\right) + \mu_{\Phi}\left(\frac{1}{2} \|\textbf{c}_{t}-\textbf{c}_{p}\|^2\right).\\
\end{aligned}
\end{equation} 
The landmark loss $L_{p}$ uses both shape and pose outputs to penalize the distance between landmarks reconstructed from ground truth or predicted shape and pose parameters (Eqs. \ref{eq:pointnorm} and \ref{eq:pca}). The loss is formulated as:
\begin{equation}
    L_{p} = \frac{1}{N} \|\textbf{p}_{t}-\textbf{p}_{p}\|^2.
    \label{eq:pointsloss}
\end{equation}
The distance loss $L_{D}$ is a combination of soft Dice loss and MSE, weighted with $\mu_{D}$:
\begin{equation}
L_{D} = \left(1 - \frac{2 \cdot \sum{S_t\cdot S_p}} {\sum{S_{t}}+\sum{S_{p}}}\right) +\mu_{D}\frac{1}{X} \|D_{t}-D_{p}\|^2
\label{eq:segloss}
\end{equation}
with $X$ the number of pixels in the image and $S$ a binary segmentation that is calculated from $D$ using $(1-sigmoid)$ as conversion function to make the loss fully differentiable:
\begin{equation}
    \label{eq:sigmoidconversion}
    S(D) = \frac{e^{-\alpha \cdot D}}{1+e^{-\alpha \cdot D}}.
\end{equation}
In this formulation, $\alpha$ affects the steepness of the transition in $S$ between 0 (outside the zero-level contour of $D$) and 1 (inside this contour).

The contour-based consistency loss $L_{C_c}$ and the overlap-based consistency loss $L_{C_o}$ combine all three outputs and are designed to enforce consistency between $\textbf{b}_p$ and $\boldsymbol{\Phi}_p$ on the one hand and $D_p$ on the other hand. $L_{C_c}$ minimizes the distance between landmarks $\textbf{p}_p$, reconstructed from predicted shape and pose parameters, and the contours of the predicted semantic segmentation, which are represented by the zero-level set of the signed distance map. Consequently, the distance can be calculated by interpolating $D_p$ at landmark positions $\textbf{p}_p$:  
\begin{equation}
    L_{C_c} = \frac{1}{N}\|D_p(\textbf{p}_p)\|^2.
    \label{eq:iploss}
\end{equation}
$L_{C_o}$ intends to maximize the overlap between the segmentation $S_p$ derived from the predicted distance map $D_p$ and the segmentation $S^{b,\Phi}_p$ corresponding to the contours derived from predicted shape and pose. $S^{b,\Phi}_p$ can be obtained by transforming a segmentation map corresponding to the mean shape $\overline{\textbf{s}}$ of the model, denoted as $\overline{S}$, using a thin-plate-spline deformation field mapping $\overline{\textbf{s}}$ onto $\textbf{p}_p$. However, since this transformation is applied during CNN training and hence requires calculation of gradients with respect to $\textbf{b}_p$, the inverse transformation mapping $\textbf{p}_p$ onto $\overline{\textbf{s}}$ and warping $D_p$ onto $D_p^T$ in the space of $\overline{S}$ is calculated instead. $S_p^T$ is subsequently obtained from $D_p^T$ using Eq. \ref{eq:sigmoidconversion}. Consequently, $L_{C_o}$ is implemented as the soft Dice loss between $\overline{S}$ and $S_p^T$:
\begin{equation}
L_{C_o} = 1 - \frac{2 \cdot \sum{\overline{S}\cdot S^{T}_p}}{\sum{\overline{S}}+\sum{S^{T}_p}}.
\end{equation}
The total loss function is a weighted sum of $L_{b}$, $L_{\Phi}$, $L_{p}$, $L_{D}$, $L_{C_c}$ and $L_{C_o}$:
\begin{equation}
    L = \gamma_{b} L_{b} + \gamma_{\Phi} L_{\Phi} + \gamma_{p} L_{p} + \gamma_{D} L_{D} + \gamma_{C_c} L_{C_c} + \gamma_{C_o} L_{C_o}.
\label{eq:totalloss}
\end{equation}

\subsection{Data} 
\label{sec:data}
For construction, training and validation of our models, three different cardiac MRI datasets were used: one in-house clinical dataset and two public datasets (ACDC (\cite{Bernard2018}) and LVQuan19 (\url{https://lvquan19.github.io/})).

The in-house dataset, referred to as 'IH' in the remainder of the text, consists of cine images of 75 subjects suffering from a wide range of cardiac pathologies. The subjects were scanned on a 1.5T MR scanner (Ingenia, Philips Healthcare, Best, The Netherlands), with a 32-channel phased array receiver coil setup. The images had a field of view (FOV) of 350$\times$350$mm^2$, acquisition matrix of 354$\times$354, acquisition resolution of 1.70$\times$1.70$mm^2$, reconstructed pixel size of\linebreak 0.99$\times$0.99$mm^2$ and slice thickness of 8$mm$. The cine scans consist of 30 time points and 10 to 18 SA slices from apex to base. The endo- and epicardium in end-diastole (ED) and end-systole (ES) images were manually delineated by a clinical expert (JB), resulting in a total of 1533 delineated 2D SA images. Additionally, we indicated the RV attachment points to allow calculation of $\theta$, as explained in Section \ref{sec:details}.
% repetition time, flip angle, sensitivity encoding factor

The second dataset was originally created for the Automated Cardiac Diagnosis Challenge (ACDC) held at STACOM'17\linebreak (\cite{Bernard2018}) and comprises cine scans of 100 subjects in five different disease groups. The images have varying FOV and acquisition matrix, and reconstructed pixel size between 0.72$\times$0.72$mm^2$ and 1.92$\times$1.92$mm^2$. More details about the dataset can be found in \cite{Bernard2018}. Ground truth segmentations for LV, myocardium and RV represented as images are available for ED and ES. This resulted in a total of 1542 delineated 2D SA images. From these segmentations, we extracted smooth endo- and epicardial and RV contours using a Gaussian smoothing kernel with standard deviation of 2. RV attachment points were automatically determined as the extremities of the connection of ground truth segmentations of RV and myocardium. 

The third dataset was obtained from the Left Ventricular Quantification Challenge held at STACOM'19 (LVQuan19,\linebreak \url{https://lvquan19.github.io/}). This dataset contains mid-cavity SA slices for the complete cardiac cycle of 56 patients with a wide variety of image size, pixel size and image appearance. Ground truth segmentations of LV and myocardium represented as images and ground truth values for 11 cardiac measures: LV and myocardial area, three LV dimensions and six regional wall thicknesses, are provided for all 20 time points comprising a total of 1120 delineated 2D SA images. We extracted endo- and epicardial contours from the ground truth segmentations using an identical approach as for the ACDC dataset and manually indicated the RV attachment points as was done for the IH dataset.

\subsection{Implementation details}
\label{sec:details}
\begin{figure}[tb]
    \centering
    \includegraphics[width=\linewidth]{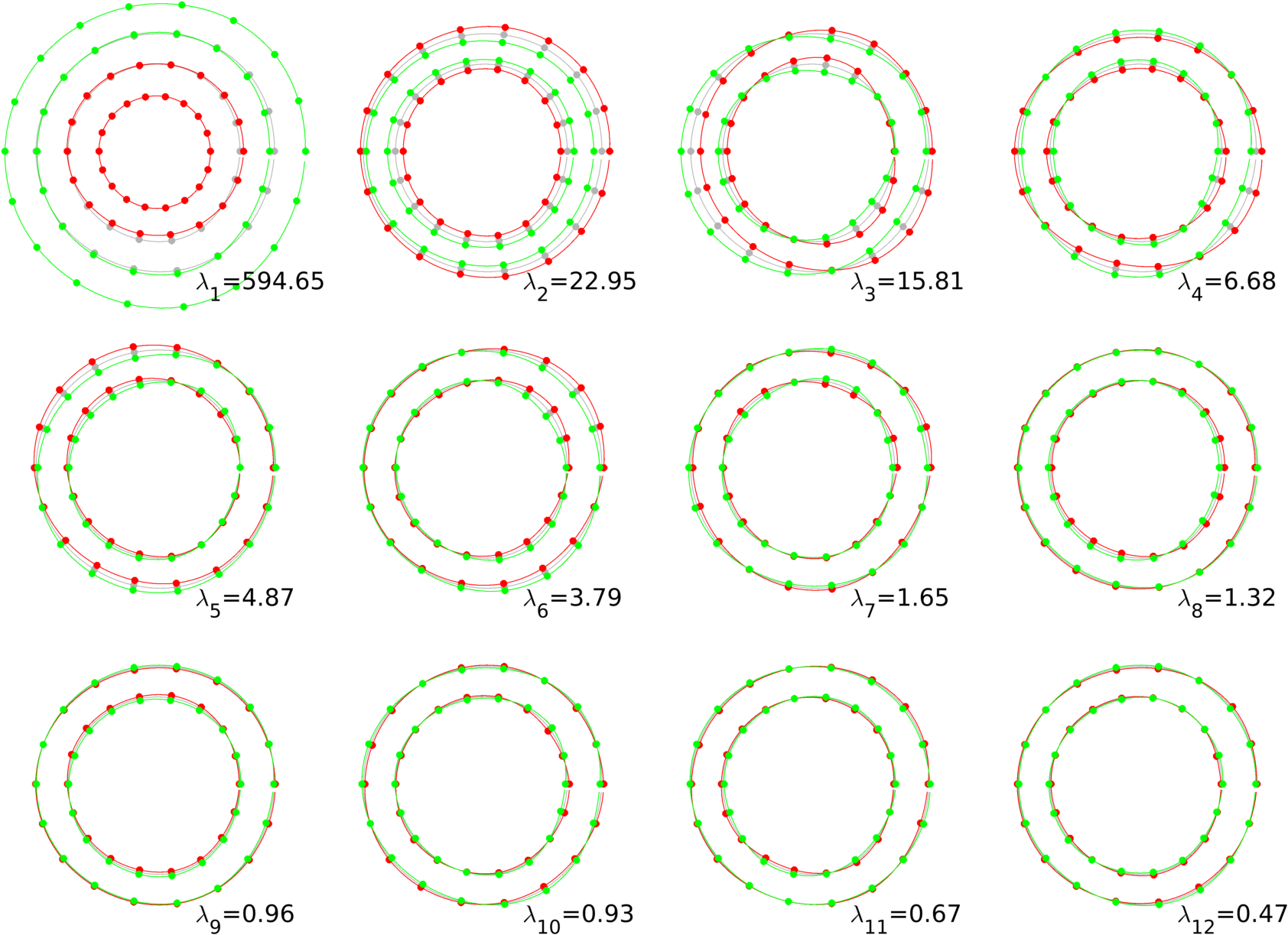}
    \caption{Typical shape variations associated with the first 12 eigenmodes and corresponding eigenvalues. $\overline{\textbf{s}} - \sqrt{\lambda_m}\cdot\textbf{v}_m$ (green), $\overline{\textbf{s}}$ (gray) and $\overline{\textbf{s}} + \sqrt{\lambda_m}\cdot\textbf{v}_m$ (red) are shown.}
    \label{fig:shapemodel}
\end{figure}

Endo- and epicardium are each represented by 18 landmarks ($N$=36) and $\theta$ is defined as the orientation of the bisector of the angle formed by the center of LV and two RV attachment points. For construction of the shape model as well as for training the CNN, 5-fold cross validation was used. For each fold in each dataset, a separate shape model was constructed using the four remaining folds. All images of a patient were assigned to the same fold. The network predicts the first $M=$ 12 shape coefficients, representing between 99.42$\%$ and 99.70$\%$ of shape variation. Shape variations associated with the first 12 eigenmodes for the first fold of dataset IH are shown in Fig.~\ref{fig:shapemodel}.

All images of IH, ACDC and LVQuan19 were resampled to a pixel size of 2$mm$x2$mm$ and image size of 128x128. The intensities were clipped at 1$\%$ and 99$\%$ before normalizing by subtracting image mean and dividing by image standard deviation. Pose parameters $\theta$, $c_x$ and $c_y$ are normalized using mean and standard deviation of the training set. Shape parameters are centered around zero with standard deviation of 1 by construction and were consequently not modified. 

The encoder part of the CNN is composed of residual layers alternated with pooling layers. Analogously, the decoder consists of residual layers alternated with upsampling layers. The parameter regression pathway contains additional convolutional layers as shown in Fig.~\ref{fig:network}. In our experiments, $A$ is equal to 8. Every convolutional layer, except for the final layer in every output, is followed by batch normalization and a parameterized rectified linear unit. Same padding is used for all convolutional layers in the encoder-decoder.

The implementation of the interpolation operation used to calculate $L_{C_c}$ is similar to interpolation layers used in CNNs for image registration (e.g. \cite{DeVos2019}) or Spatial Transformer Networks (\cite{Jaderberg2015}). Linear interpolation was used to calculate distance values at non-integer pixel positions. By avoiding the use of non-differentiable functions for implementation of the interpolation, a gradient towards the regression pathway as well as towards the semantic segmentation pathway is generated.

The warped, predicted distance map $D_p^T$ is obtained by warping $D_p$ onto the space of $\overline{S}$. For this, a spatial transformation layer with linear interpolation, as discussed above, was used. After transformation, the pixels in $D_p^T$ do no longer represent the distance to the zero-level contour in the new image space. To correct for large scaling differences that will impact the transformation to $S^T_p$, we perform a scaling $a \cdot D_p^T$ before applying Eq.~\ref{eq:sigmoidconversion}. Scaling factor $a$ is calculated as the ratio between the average distance of $\overline{\textbf{s}}$ to the origin and the average distance of $\textbf{p}_p$ to the LV center $\textbf{c}_p$. 

\subsection{Training}
\label{sec:training}
During training, online data augmentation is applied\linebreak by adapting pose and shape parameters. Position and orientation offsets are sampled from uniform distributions between $[-40,40] mm$ and $[-\pi/2,\pi/2] rad$, respectively. Additionally, shape coefficients were adapted as $b_{m,aug} = b_m + \delta b$, where $\delta b$ is sampled from a uniform distribution between -1 and +1 (i.e. one standard deviation). The input images are warped accordingly based on a thin-plate-spline deformation between the original and augmented landmark locations while the distance maps are recreated from the augmented landmarks as described in Section \ref{sec:shapemodel}. Furthermore, Gaussian noise with standard deviation between 0 and 0.1 is online added to the MR images during training.

The weights used in Eq.~\ref{eq:totalloss} are $\gamma_{b} = 1$, $\gamma_{\Phi} = 1$, $\gamma_{p} = 1$, $\gamma_{D} = 100$, $\gamma_{C_c}=1$, $\gamma_{C_o}=10$, $\mu_{\Phi} = 1$ and $\mu_{D} = 0.1$. Parameter $\alpha$ in Eq.~\ref{eq:sigmoidconversion} is set to 5, resulting in a value of $S = 0.0067$ and $S=0.9933$ at a distance of one pixel outside and inside the zero-level set of $D$, respectively. 
The networks are trained with Adam optimizer using 
the adaptive learning rate scheme [2e-3, 1e-3, 5e-4, 2e-4] for [2000, 200, 200, 100] epochs 
with a batch size of 32. For experiments were pose and shape data augmentation was omitted, the networks were trained for only half the number of epochs for every learning rate due to faster convergence. 
If the training started from a pretrained network (Section \ref{sec:consistency}), the first 2000 epochs were used for pretraining and the final 500 epochs for finetuning. All training parameters were tuned on fold 1 of dataset IH and kept the same for all other experiments.

The CNNs were implemented in Keras using TensorFlow and trained on a NVIDIA GeForce RTX 2080 Ti GPU.

\section{Experiments}
\begin{table}[tb]
     \caption{Overview of the performed experiments. The type of geometric data augmentation (A) is indicated: None (-), pose ($p$) or model-based shape augmentation ($s$), as well as the weights of Eq.~\ref{eq:totalloss} that were set to zero (-). The last three experiments were optimized starting from a pretrained network using setup 'RS-A$_{ps}$-$L_p$'.}
    \label{tab:overviewExperiments}
    \centering
    \scriptsize
    \begin{tabular}{|l|c|cccccc|cc|}
    \hline
        & A & $\gamma_{b}$& $\gamma_{\Phi}$& $\gamma_{p}$ & $\gamma_{D}$&  $\gamma_{C_c}$ & $\gamma_{C_o}$ & $\mu_{\Phi}$& $\mu_D$ \\
        \hline
        B & - & - & - & - & + & - & - & -& - \\
        S& - & - & - & - & + & - & - & -& + \\
        S-A$_p$&$p$ & - & - & - & + & - & - & -& + \\
        S-A$_{ps}$& $ps$ & - & - & - & + & - & - & -& + \\
        \hline
    	R & - & + & + & - & -& -& -& +& -\\
		R-A$_{p}$& $p$ & + & + & - & -& -& -& +& -\\
		R-A$_{ps}$ & $ps$ & + & + & - & -& -& -& +& -\\
		R-A$_{ps}$-$L_{p}$& $ps$ & + & + & + & -& -& -& +& -\\
		\hline
		RS-A$_{ps}$-$L_{p}$& & & & & & & & & \\
        \multicolumn{1}{|c|}{=} & $ps$ & + & + & + & +& -& -& +& +\\
		\multicolumn{1}{|c|}{baseline}& & & & & & & & & \\
		\hline
		\hline
		$L_{C_c}$& $ps$& - & + & - & +& +& -& -& +\\
		$L_{C_o}$& $ps$& - & + & - & +& -& +& -& +\\
		$L_{C_c}$+$L_{C_o}$& $ps$& - & + & - & +& +& +& -& +\\
		\hline
    \end{tabular}
\end{table}
We performed different experiments to validate our method and the added value of every choice. First, we evaluated and optimized the performance of CNNs trained for either semantic segmentation or for pose and shape regression separately. Second, a CNN was trained to jointly predict a distance map and shape and pose parameters. Third, the losses $L_{C_c}$ and $L_{C_o}$ which are intended to explicitly enforce consistency between predicted distance map and predicted shape and pose parameters, were introduced. An overview of all different experiments is given in Table \ref{tab:overviewExperiments}. 
Finally, we compared the results of our~proposed method with a state-of-the-art segmentation network nnUNet (\cite{Isensee}).

\subsection{Semantic segmentation using distance maps}
\label{sec:seg}
The first set of experiments is intended to evaluate the performance of distance map prediction for semantic segmentation. We start with a conventional semantic segmentation to predict a binary ('B') segmentation map. To this end, the final activation layer in the ResU block of Fig.~\ref{fig:network} was converted to a sigmoid activation function and the soft Dice loss was used for training. This approach was compared to our proposed distance map prediction using combined MSE and soft Dice loss ('S'). Additionally, we compare the impact of conventional pose data augmentation ('S-A$_p$') with our proposed model-based shape data augmentation ('S-A$_{ps}$'). 

\subsection{Prediction of shape and pose}
\label{sec:reg}
In these experiments, the capability of our CNN to predict shape and pose parameters without simultaneous semantic segmentation is investigated. The benefit of our model-based shape data augmentation is investigated in three different experiments: 1) training without pose or shape data augmentation~('R'), 2)~with conventional pose data augmentation ('R-A$_p$') and 3)~with the proposed model-based shape data augmentation ('R-A$_{ps}$'). Furthermore, we evaluate if the loss $L_p$ on landmarks constructed from $\textbf{b}$ and $\boldsymbol{\Phi}$ improves the performance ('R-A$_{ps}$-$L_p$'). Compared to the more abstract shape coefficients $\textbf{b}$, the landmarks $\textbf{p}$ can be directly linked to the MR images.

\subsection{Shape prediction with joint semantic segmentation}
\label{sec:segreg}
This experiment jointly predicts distance map and shape and pose parameters ('RS-A$_{ps}$-$L_p$'). The results will be compared to the previously described experiments predicting distance map or shape and pose parameters separately. 

\subsection{Consistent prediction with $L_{C_c}$ and $L_{C_o}$}
\label{sec:consistency}
Finally, the impact of the losses $L_{C_c}$ and $L_{C_o}$, designed to enforce consistency between $\textbf{b}_p$ and $\boldsymbol{\Phi_p}$ on the one hand and $D_p$ on the other hand, was investigated. First, we defined our baseline to be experiment RS-A$_{ps}$-$L_{p}$. Second, we used the consistency losses $L_{C_c}$ and/or $L_{C_o}$ for further training starting from a pretrained network obtained with setup RS-A$_{ps}$-$L_{p}$ as explained in Section \ref{sec:training}. Since the CNNs are pretrained, we assume a good initialization for $\textbf{b}_p$, $\boldsymbol{\Phi_p}$ and $D_p$. Furthermore, since we value a good consistency more than the exact correspondence of shape parameters $\textbf{b}$, especially for the higher modes, and since the midpoint $\textbf{c}$ is fully determined by the consistency loss, we set $\gamma_b$, $\gamma_p$ and $\mu_{\Phi}$ to zero. This way, the effect of $L_{C_c}$ and $L_{C_o}$ can be better studied. $\gamma_{\Phi}$ remained 1 because $L_{C_c}$ and $L_{C_o}$ are not sensitive to orientation. $\gamma_{D}$ was retained to prevent divergence from the ground truth. 

\subsection{Comparison with state-of-the-art}
\label{sec:comparision}
We compare the performance of our proposed method with nnUNet (\cite{Isensee2021}), which has won multiple segmentation challenges. A main difference in architecture with our method is the use of skip connections. Hence, we trained an additional CNN similar to our baseline method but with skip connections added between encoder and decoder to study the effect of this difference. We additionally trained multi-class segmentation networks, segmenting LV cavity, RV cavity and myocardium, using the publicly available nnUNet code (\url{https://github.com/MIC-DKFZ/nnUNet}) out-of-the-box. \linebreak First, we trained a 2D CNN on data identically preprocessed as in our method. Next, we trained both 2D and 3D CNNs on the raw ACDC challenge data. The ensemble of 2D and 3D predictions constitutes the nnUNet method proposed by \cite{Isensee2021}.

\subsection{Evaluation}
All experiments described in Sections \ref{sec:seg}, \ref{sec:reg} and \ref{sec:segreg} were performed in a 5-fold cross validation on dataset IH. We first evaluated the segmentation accuracy of the two predicted representations of the myocardium: the segmentation from the predicted distance map, referred to as 'Map' in the results, and the segmentation obtained from predicted shape and pose parameters as described in Section \ref{sec:shapemodel}, referred to as 'Contour'. For both representations, Dice similarity coefficient (DSC), mean boundary error (MBE) and Hausdorff distance (HD) of the myocardium were calculated. Additionally, for the 'Map' predictions, the number of cases where the predicted segmentation map was empty ($n_\phi$) and the number of predictions considered to have an unrealistic shape ($n_{err}$) were reported. Unrealistic shapes include the cases with no myocardial segmentation (=$n_\phi$), cases with no LV cavity, cases with an open myocardium or segmentations with more than one connected component.

Next, the performance of shape and pose prediction was validated. The position and orientation errors were respectively defined as $\Delta \textbf{c} = \|\textbf{c}_t-\textbf{c}_p\|$ and $\Delta\theta=|\theta_{t}-\theta_{p}|$. The impact of errors in the prediction of shape coefficients $\textbf{b}$ on landmarks $\textbf{p}$ was validated by calculating the Euclidean distance between ground truth landmarks and landmarks reconstructed using predicted shape coefficients and ground truth pose parameters ($\Delta\textbf{p}_b$). Additionally, influence of every shape coefficient $b_m$ on this distance error was calculated by reconstructing the landmarks with an increasing number of coefficients, i.e. first only $\{b_1\}$ is used, then $\{b_1,b_2
\}$, then $\{b_1,b_2,b_3\}$... Furthermore, mean absolute error (MAE) and Pearson correlation coefficient $\rho$ for every shape coefficient were obtained.

\begin{figure}
    \centering
    \includegraphics[width=0.6\linewidth]{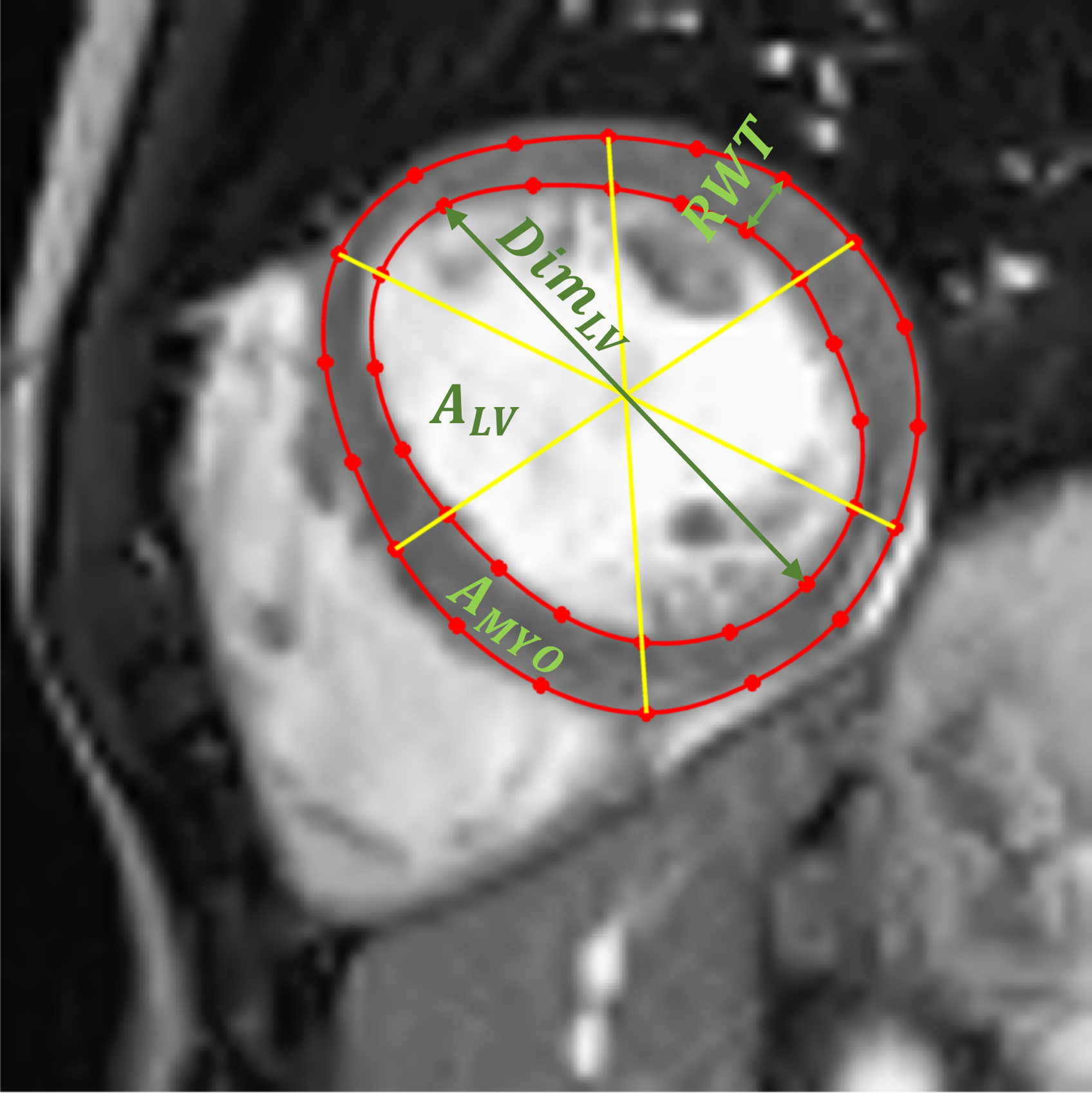}
    \caption{Definition of LV area ($A_{LV}$), myocardial area ($A_{MYO}$), LV dimension ($Dim_{LV}$) and regional wall thickness ($RWT$).}
    \label{fig:physicalmetrics}
\end{figure}

For validation of the consistency losses $L_{C_c}$ and $L_{C_o}$, a 5-fold cross validation on all three datasets was performed. Similar to the previous experiments, DSC, MBE and HD were calculated between 'Contour' and ground truth, between 'Map' and ground truth and between 'Contour' and 'Map'. Furthermore, the LV parameters used in the LVQuan18 (\url{https://lvquan18.github.io/}) and LVQuan19 (\url{https://lvquan19.github.io/}) challenges including LV area ($A_{LV}$), myocardial area ($A_{MYO}$), LV dimensions in three different orientations \linebreak ($Dim_{LV}$) and regional wall thickness ($RWT$) for six cardiac segments, as shown in Fig. \ref{fig:physicalmetrics}, were calculated. These geometric parameters are useful to assess pathological shape variations and to monitor disease progression (\cite{Bogaert2012}). Hypertrophic cardiomyopathy is for example characterized by an increase in myocardial area. The associated thickening of the myocardium can be either symmetric or asymmetric, which is expressed by $RWT$. Dilated cardiomyopathy results in an increase in LV area and consequently in LV dimensions. Furthermore, LV area relates to LV volume and hence ejection fraction, which is an important parameter of cardiac function (\cite{Bogaert2012}). The LV parameters can be straightforwardly calculated from the 'Contour' output. $Dim_{LV}$ and $RWT$ were obtained by calculating the distance between two corresponding landmarks and the resulting distances in one segment were averaged. For $A_{LV}$ and $A_{MYO}$, the binary reconstruction introduced to calculate DSC was used. Note that these metrics are independent of pose $\boldsymbol{\Phi}$, such that only shape parameters $\textbf{b}$ and the underlying model are required. The ground truth to which these values were compared, was constructed identically. For these four shape properties, MAE and $\rho$ were calculated. Additionally, we also calculated the consistency of these metrics when extracted from the 'Contour' or 'Map' outputs. To obtain $Dim_{LV}$ and $RWT$ from the 'Map' output, the predicted orientation $\theta_p$ was additionally required. Landmarks were sampled on both endo- and epicardium at equiangular distances relative to $\theta_p$, in the same way as was done to construct the shape model. In case of an erroneous, unrealistic segmentation, previously defined as $n_{err}$, the four shape properties could not be calculated.
 
Besides myocardial DSC, MBE and HD, we also calculated the four LV parameters in the comparison with the state-of-the-art which was performed on dataset ACDC. For the multi-class nnUNet segmentations, the cardiac orientation required to calculate $RWT$ and $Dim_{LV}$ was automatically obtained by extracting RV attachment points as the extremities of the connection of predicted RV and myocardial segmentations, in the same way as was done for the ground truth. This approach does not allow calculation of orientation if the prediction did not contain RV, or if RV and myocardial predictions were not connected. The total number of cases for which $RWT$ and $Dim_{LV}$ could not be calculated due to an unrealistic segmentation of the myocardium ($n_{err}$) or due to a failure in the calculation of the cardiac orientation was reported as $n_{err,\theta}$.

\subsection{Statistical testing} 
Statistically significant differences were assessed with non-parametric bootstrapping, making no assumptions on the distribution of the results (\cite{Bakas2018}). In this method, all images were first individually ranked according to their metric values before obtaining a single rank for every experiment by averaging. To calculate the statistical significance of the difference between a pair of experiments, the ranks of all images were randomly permuted 100,000 times. For every permutation, the difference in average rank was calculated. The p-value was obtained as the frequency this difference exceeds the true difference in ranks. Results were considered statistically significant if the p-value was below 5$\%$.

\section{Results}
\begin{table*}[tb]
\centering
	\caption{Mean and standard deviation for DSC, MBE and HD calculated between segmentations obtained from predicted shape and pose parameters and ground truth ('Contour-Gt'), and between segmentations obtained from distance map prediction and ground truth ('Map-Gt'). Additionally, the number of cases where the predicted segmentation map was empty ($n_\phi$) and the number of predictions with unrealistic shapes ($n_{err}$) are reported. Statistically significant best values are indicated in bold. Statistically significant improvement with respect to the previous row is indicated with '$^+$'. The cases reported in $n_\phi$ were not used for calculation of MBE and HD.}
	\label{tab:Seg}
	\scriptsize
	\begin{tabular}{|l|l|l|l|l|l|l|l|l|}
		\hline
		& \multicolumn{3}{c|}{Contour-Gt} & \multicolumn{5}{c|}{Map-Gt}\\
		IH & \multicolumn{1}{c|}{DSC [$\%$]} & \multicolumn{1}{c|}{MBE [pix]} & \multicolumn{1}{c|}{HD [pix]} & \multicolumn{1}{c|}{DSC [$\%$]} & \multicolumn{1}{c|}{MBE [pix]} & \multicolumn{1}{c|}{HD [pix]} &$n_\phi$ & $n_{err}$\\
		\hline
		B &&&&83.5 $\pm$ 11.6& 0.72 $\pm$ 0.50& 3.07 $\pm$ 5.92&1&108\\
		S &&&&83.2 $\pm$ 11.1& 0.70 $\pm$ 0.33& 2.11 $\pm$ 2.46$^+$&1&44 \\
		S-A$_{p}$ &&&&\textbf{85.5 $\pm$ 9.4}$^+$& \textbf{0.60 $\pm$ 0.23}$^+$& \textbf{1.71 $\pm$ 1.58}$^+$ &1&20\\
		S-A$_{ps}$ &&&& \textbf{85.7 $\pm$ 8.9}& \textbf{0.60 $\pm$ 0.23}& \textbf{1.72 $\pm$ 1.62} &0&29\\
		\hline
		R &42.6 $\pm$ 22.5& 2.41 $\pm$ 2.18& 6.64 $\pm$ 5.07& && && \\
		R-A$_{p}$& 72.9 $\pm$ 16.4$^+$& 0.97 $\pm$ 0.43$^+$& 2.63 $\pm$ 1.75$^+$&&&&& \\
		R-A$_{ps}$ & 78.1 $\pm$ 13.5$^+$& 0.82 $\pm$ 0.34$^+$& 2.19 $\pm$ 1.58$^+$&&&&&\\
		R-A$_{ps}$-$L_p$ & 82.0 $\pm$ 10.3$^+$& 0.72 $\pm$ 0.28$^+$& 1.97 $\pm$ 1.51$^+$&&&&&\\
		\hline
		RS-A$_{ps}$-$L_{p}$&\textbf{83.3 $\pm$ 9.8}$^+$& \textbf{0.68 $\pm$ 0.27}$^+$& \textbf{1.89 $\pm$ 1.57}$^+$& 85.6 $\pm$ 8.7& 0.60 $\pm$ 0.24& \textbf{1.73 $\pm$ 1.69}&0&32 \\
		\hline
	\end{tabular}
\end{table*}

\begin{figure*}[tb]
    \centering
    \includegraphics[width = 0.9\linewidth]{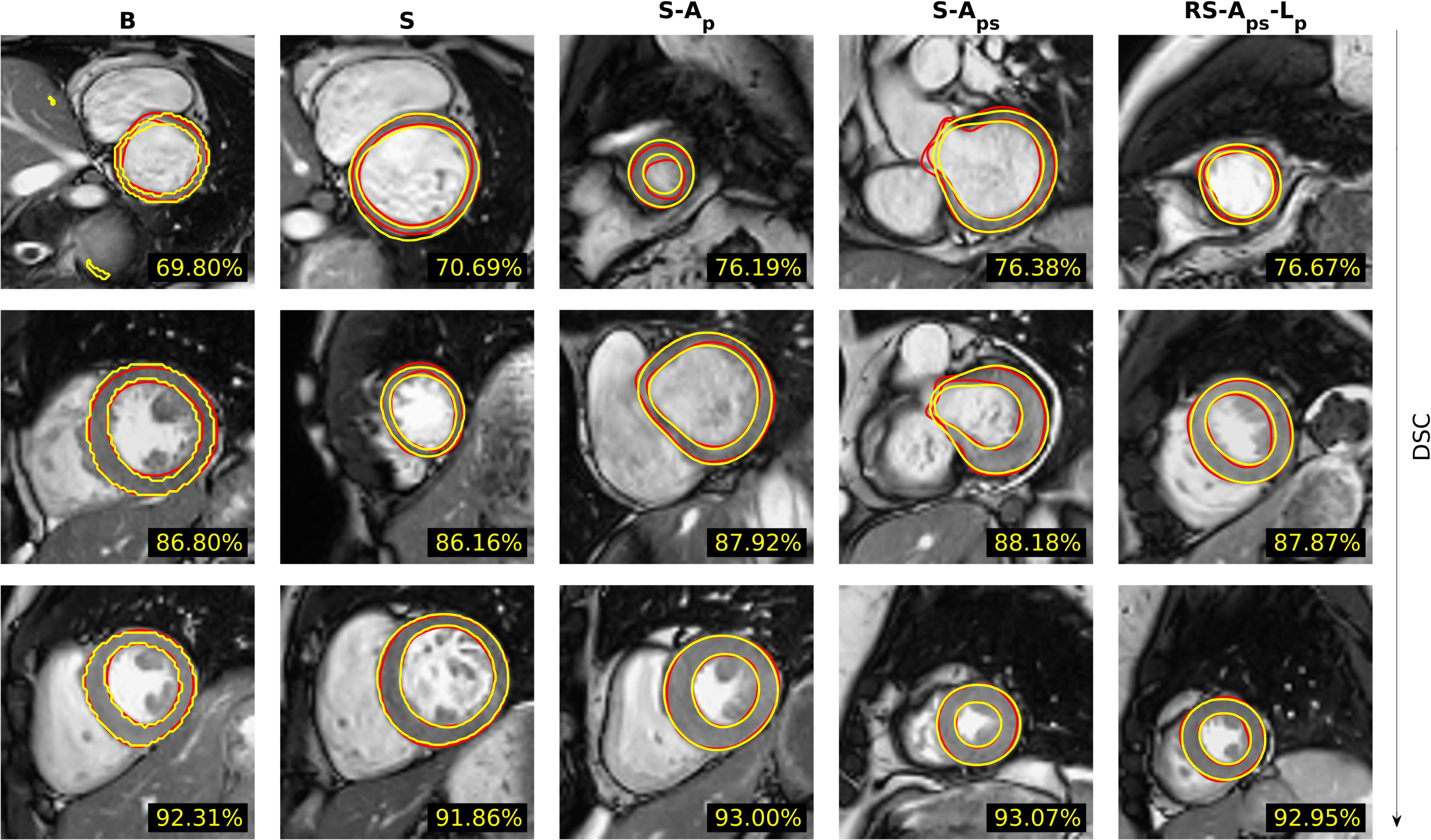}
    \caption{Segmentation results obtained from distance map prediction ('Map', yellow) and ground truth segmentations (red). The cases were selected as 10$\%$, 50$\%$ and 90$\%$ (top to bottom) of the 'Map' myocardial DSC, which is shown in yellow on the bottom right of every image. Cropped images are used for visualization purpose.}
    \label{fig:segresults_map}
\end{figure*}

Table \ref{tab:Seg} shows the segmentation results of semantic segmentation and shape and pose prediction considered as separate tasks or trained jointly. When focusing on semantic segmentation, we observe that distance map regression (S) achieved similar DSC and MBE compared to binary segmentation (B), while HD was significantly improved. Furthermore, distance map regression also decreased the number of cases with unrealistic shapes from 108 (=$7.0\%$) to 44 (=$2.9\%$), which is a reduction with 59$\%$. By adding conventional data augmentation (S-A$_{p}$), all metrics significantly improved compared to (S). Introducing model-guided shape data augmentation (S-A$_{ps}$) did not further enhance the results. Example segmentations for the different experiments, selected as 10$\%$, 50$\%$ and 90$\%$ of myocardial DSC, are shown in Fig.~\ref{fig:segresults_map}. For the 10$\%$ binary segmentation (B) example, some disconnected components outside the heart are observed. This error arises because, in contrast to distance-based loss functions, the soft Dice loss is not sensitive to such small components. Furthermore, contour extraction from distance maps (S) resulted in smoother contours compared to the binary segmentation (B).

\begin{figure*}[tb]
    \centering
    \includegraphics[width = 0.9\linewidth]{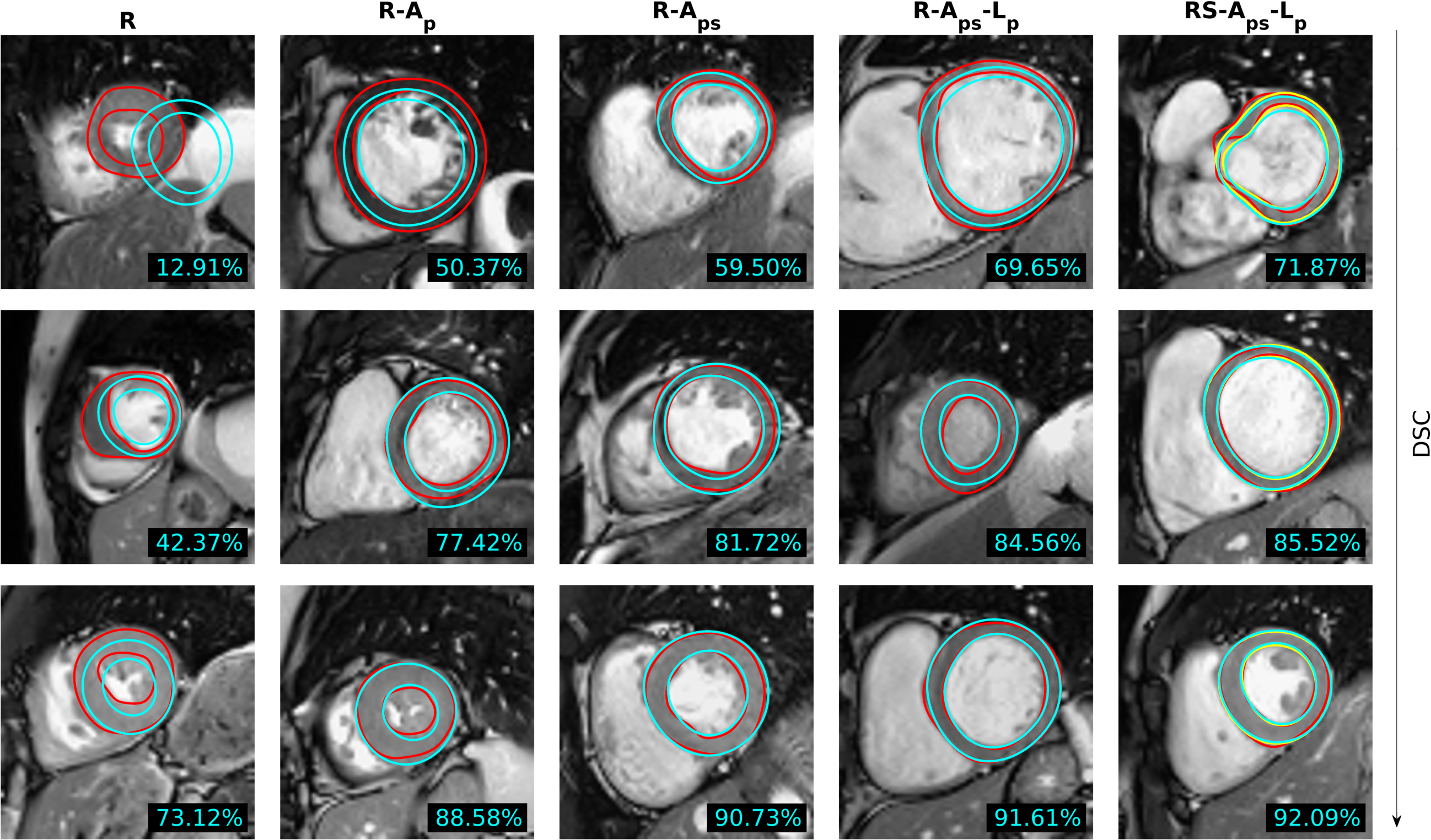}
    \caption{Segmentation results obtained from predicted shape and pose parameters ('Contour', cyan) and ground truth segmentations (red). For setup 'RS-A$_{ps}$-$L_p$', the segmentation result obtained from predicted distance map ('Map', yellow) is also shown. The cases were selected as 10$\%$, 50$\%$ and 90$\%$ (top to bottom) of the 'Contour' myocardial DSC, which is shown in cyan on the bottom right of every image. Cropped images are used for visualization purpose.}
    \label{fig:segresults_contour}
\end{figure*}

The segmentation results obtained for shape and pose prediction ('Contour') shown in Table \ref{tab:Seg} indicate a poor segmentation if no augmentation is applied (R), with an average DSC of only 42.6$\%$. This is illustrated by the example segmentations in Fig.~\ref{fig:segresults_contour} where there is little agreement between ground truth and prediction, especially for the examples at 10$\%$ and 50$\%$ of myocardial 'Contour' DSC. The segmentation significantly improved by applying conventional pose data augmentation (R-A$_p$). Furthermore, in contrast to semantic segmentation, model-guided shape data augmentation (R-A$_{ps}$) did further improve DSC, MBE and HD. Training additionally on landmark loss $L_{p}$ instead of using separate shape ($L_{b}$) and pose ($L_{\Phi}$) losses (R-A$_{ps}$-$L_p$) resulted again in significantly better predictions. 

The joint prediction of distance map, shape and pose (RS-A$_{ps}$-$L_p$) significantly improved DSC, MBE and HD calculated from predicted shape and pose parameters. While the addition of the semantic segmentation pathway to guide the regression was thus successful, we note that the metrics obtained for the 'Contour' segmentation remained significantly worse compared to the 'Map' segmentation for experiment (RS-A$_{ps}$-$L_p$). % STATEMENT HAS BEEN CONFIRMED BY STATISTICAL TEST

Table \ref{tab:Reg} reports the position error ($\Delta \textbf{c}$), orientation error ($\Delta \theta$) and the impact of shape prediction on landmark error ($\Delta \textbf{p}_b$) for the different experiments. The position and orientation estimations were significantly improved by using conventional data augmentation (R-A$_p$), by using the model-guided shape data augmentation (R-A$_{ps}$) and by using the landmark loss $L_p$ (R-A$_{ps}$-$L_p$) but did not benefit from joint semantic segmentation. The landmark error caused by errors in predicted shape coefficients only ($\Delta \textbf{p}_b$) showed significant improvement for every alteration. 
\begin{table}%[tb]
\centering
	\caption{Mean and standard deviation for position error ($\Delta \textbf{c}$), orientation error ($\Delta \theta$) and the impact of shape prediction on landmark error ($\Delta \textbf{p}_b$). Best values are indicated in bold. Statistically significant improvement with respect to the previous row is indicated with '$^+$'.
	}
	\label{tab:Reg}
	\scriptsize
	\begin{tabular}{|l|l|l|l|}
		\hline
		IH& \multicolumn{1}{c|}{$\Delta \textbf{c}$ [$pix$]} & \multicolumn{1}{c|}{$\Delta\theta$ [$^\circ$]} & \multicolumn{1}{c|}{$\Delta \textbf{p}_b$ [$pix$]}\\
		\hline
		R & 4.32 $\pm$ 4.46& 10.07 $\pm$ 9.77& 1.75 $\pm$ 1.29\\
		
		R-A$_{p}$&1.04 $\pm$ 0.81$^+$& 7.44 $\pm$ 7.03$^+$& 0.96 $\pm$ 0.60$^+$ \\
	
		R-A$_{ps}$ & 0.93 $\pm$ 0.67$^+$& 6.20 $\pm$ 6.11$^+$& 0.80 $\pm$ 0.49$^+$\\
		R-A$_{ps}$-$L_{p}$&\textbf{0.82 $\pm$ 0.64}$^+$& \textbf{4.59 $\pm$ 5.43}$^+$& 0.69 $\pm$ 0.42$^+$\\
		\hline
		RS-A$_{ps}$-$L_{p}$&\textbf{0.82 $\pm$ 0.65}& 5.29 $\pm$ 6.23& \textbf{0.66 $\pm$ 0.41}$^+$\\ 
	    \hline
	\end{tabular}
\end{table}

\begin{figure*}%[tb]
    \centering
    \includegraphics[width=\linewidth]{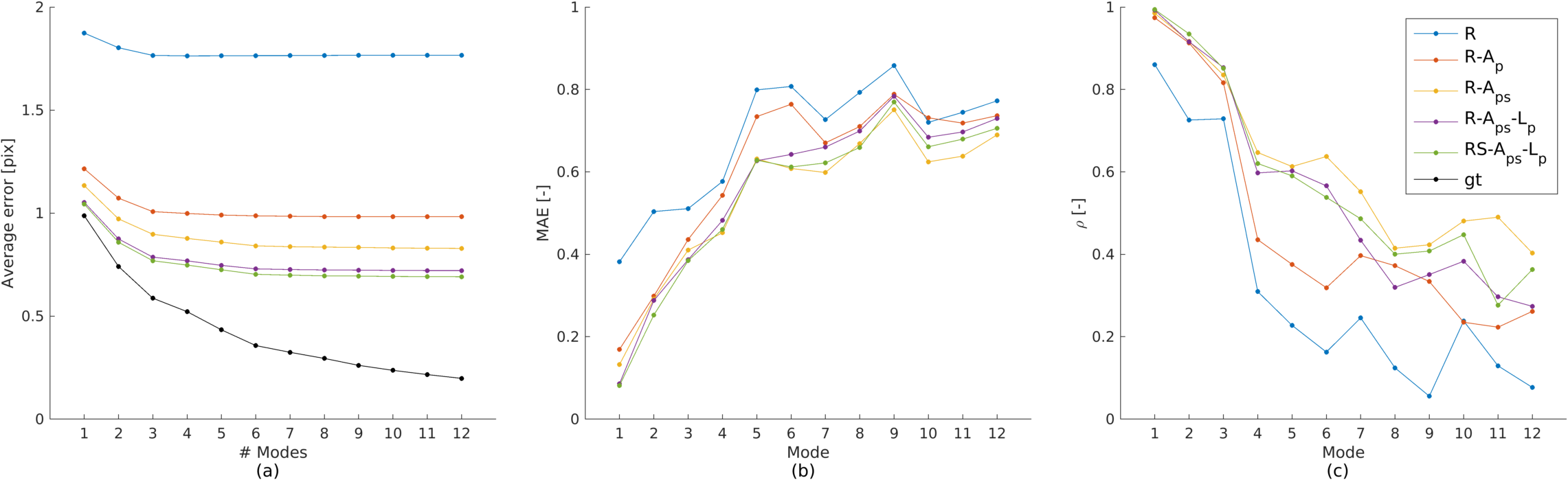}
    \caption{Average distance between ground truth landmarks and landmarks reconstructed using a limited number of predicted coefficients $b_m$ and ground truth pose parameters (a), MAE (b) and correlation (c) for every $b_m$.}
    \label{fig:shape}
\end{figure*}

The contribution of every shape coefficient to this landmark error $\Delta \textbf{p}_b$ is shown in Fig.~\ref{fig:shape}a where the error between ground truth landmarks and landmarks reconstructed using a limited number of ground truth or predicted shape coefficients is shown. While the error consistently decreases when using more ground truth coefficients, this decrease is less pronounced for the predicted coefficients. After a few modes, the curves flatten indicating that higher coefficients are more difficult to predict accurately. Fig.~\ref{fig:shape}b and c show MAE and correlation of individual shape coefficients, respectively. While the first shape coefficient, accounting for the largest shape variation, is well predicted with a MAE of 0.08 and $\rho$ of 0.99 for RS-A$_{ps}$-$L_{p}$, MAE increases and $\rho$ decreases for higher coefficients, reaching a MAE of 0.71 and $\rho$ of only 0.36 for the twelfth coefficient. 

Table \ref{tab:RegSegPost} reports the segmentation results for datasets IH, ACDC and LVQuan19 using no consistency loss ('baseline'), the contour-based consistency loss $L_{C_c}$, the overlap-based consistency loss $L_{C_o}$ or both losses ($L_{C_c}$+$L_{C_o}$). For all three datasets, both losses and their combination significantly improved DSC, MBE and HD calculated from the 'Contour' outputs, except for $L_{C_c}$ in dataset ACDC. The 'Map' segmentations show overall stable results where 18/27 metrics of $L_{C_c}$, $L_{C_o}$ or $L_{C_c}$+$L_{C_o}$ did not show significant differences compared to the baseline, 2/27 metrics showed an improvement and 7/27 metrics a deterioration. All metrics, except for $L_{C_c}$ in dataset ACDC, showed a significant improvement of the consistency between both outputs. 

Although the results in Table \ref{tab:RegSegPost} show the benefits of a consistency loss, it is less clear whether $L_{C_c}$, $L_{C_o}$ or their combination is to be preferred. $L_{C_c}$ was statistically significantly best for 4/9 metrics for 'Contour-Gt', $L_{C_o}$ for 6/9 and $L_{C_c}$+$L_{C_o}$ for 7/9. For 'Map-Gt', $L_{C_c}$ was best for 6/9 metrics, $L_{C_o}$ for 3/9 and $L_{C_c}$+$L_{C_o}$ for 5/9. If the consistency between 'Contour' and 'Map' is to be preferred, $L_{C_o}$ scores best with superior results for 9/9 metrics, while this is only 2/9 for $L_{C_c}$ and $L_{C_c}$+$L_{C_o}$.

% Experiment 3, lre-3,w1 0 0 0 0 1,w1 0 0 0 0 0k1

\begin{table*}[tb]
\centering
	\caption{Mean and standard deviation for DSC, MBE and HD. Results obtained from shape and pose parameters ('Contour-Gt'), from the distance maps ('Map-Gt') or the consistency between both outputs ('Contour-Map') are reported. Best values, for 'Contour-Gt', 'Map-Gt' and 'Contour-Map' and every dataset separately, are indicated in bold. Statistically significant improvement of every metric compared to baseline ($b$), $L_{C_c}$ ($c$), $L_{C_o}$ ($o$) or $L_{C_c}$+$L_{C_o}$ (joint, $j$) is indicated.}
	\scriptsize
	\label{tab:RegSegPost}
	\begin{tabular}{|l|l|l|l|l|l|l|l|l|l|}
		\hline
		& \multicolumn{3}{c|}{Contour-Gt}& \multicolumn{3}{c|}{Map-Gt}& \multicolumn{3}{c|}{Contour-Map}\\
		&  \multicolumn{1}{c|}{DSC[$\%$]} &\multicolumn{1}{c|}{MBE [pix]}& \multicolumn{1}{c|}{HD [pix]} & \multicolumn{1}{c|}{DSC [$\%$]} & \multicolumn{1}{c|}{MBE [pix]}& \multicolumn{1}{c|}{HD [pix]} & \multicolumn{1}{c|}{DSC [$\%$]} & \multicolumn{1}{c|}{MBE [pix]}& \multicolumn{1}{c|}{HD [pix]}\\
		\hline
		\underline{IH}&&&&&&&&& \\
		baseline & 83.3 $\pm$ 9.8& 0.68 $\pm$ 0.27& 1.89 $\pm$ 1.57& 85.6 $\pm$ 8.7& \textbf{0.60 $\pm$ 0.24}$^o$& \textbf{1.73 $\pm$ 1.69}$^o$& 88.8 $\pm$ 5.6& 0.46 $\pm$ 0.15& 1.10 $\pm$ 0.67\\
		
		$L_{C_c}$& 83.9 $\pm$ 9.8$^b$& \textbf{0.66 $\pm$ 0.26}$^b$& \textbf{1.83 $\pm$ 1.51}$^b$& 85.6 $\pm$ 8.8& \textbf{0.60 $\pm$ 0.24}$^{o}$& \textbf{1.73 $\pm$ 1.64}& 90.1 $\pm$ 5.1$^b$& 0.42 $\pm$ 0.13$^{b}$& 1.02 $\pm$ 0.64$^b$\\ % IP postproc
		
		$L_{C_o}$& \textbf{84.2 $\pm$ 9.3}$^{bc}$& \textbf{0.66 $\pm$ 0.26}$^b$& \textbf{1.85 $\pm$ 1.53}$^b$& 85.7 $\pm$ 8.6& 0.60 $\pm$ 0.24& 1.73 $\pm$ 1.64 & \textbf{91.0 $\pm$ 4.6}$^{bcj}$& \textbf{0.41 $\pm$ 0.12}$^{bcj}$& \textbf{0.98 $\pm$ 0.58}$^{bc}$\\% TPS postproc
		
		$L_{C_c}$+$L_{C_o}$& \textbf{84.1 $\pm$ 9.3}$^{bc}$& \textbf{0.66 $\pm$ 0.26}$^b$& \textbf{1.83 $\pm$ 1.47}$^b$& \textbf{85.8 $\pm$ 8.6}$^{bco}$& \textbf{0.60 $\pm$ 0.23}& \textbf{1.71 $\pm$ 1.59}$^{o}$& 90.5 $\pm$ 4.9$^{bc}$& 0.41 $\pm$ 0.13$^{bc}$& \textbf{0.98 $\pm$ 0.57}$^{bc}$\\
		
	    \underline{ACDC}&&&&&&&&&\\
	    
	    baseline& 84.4 $\pm$ 8.8$^c$& 0.75 $\pm$ 0.25$^c$& 1.97 $\pm$ 0.76$^c$& 88.0 $\pm$ 6.8& \textbf{0.61 $\pm$ 0.21}& \textbf{1.61 $\pm$ 0.71}& 88.9 $\pm$ 6.4$^c$& 0.56 $\pm$ 0.20$^c$& 1.33 $\pm$ 0.56$^c$\\
        
        $L_{C_c}$& 83.8 $\pm$ 8.9& 0.77 $\pm$ 0.24& 2.01 $\pm$ 0.75& 88.0 $\pm$ 6.8 & \textbf{0.60 $\pm$ 0.20}& \textbf{1.61 $\pm$ 0.69}& 88.2 $\pm$ 6.6& 0.58 $\pm$ 0.21& 1.36 $\pm$ 0.56\\

        $L_{C_o}$&\textbf{84.9 $\pm$ 8.6}$^{bc}$& \textbf{0.73 $\pm$ 0.24}$^{bcj}$& \textbf{1.93 $\pm$ 0.75}$^{bcj}$& \textbf{88.2 $\pm$ 7.1}$^{bcj}$& \textbf{0.60 $\pm$ 0.23}& \textbf{1.61 $\pm$ 0.79}& \textbf{89.8 $\pm$ 6.0}$^{bcj}$& \textbf{0.53 $\pm$ 0.20}$^{bcj}$& \textbf{1.24 $\pm$ 0.55}$^{bcj}$\\
        
        $L_{C_c}$+$L_{C_o}$& \textbf{84.8 $\pm$ 8.5}$^{bc}$& 0.74 $\pm$ 0.23$^{bc}$& 1.94 $\pm$ 0.74$^{bc}$& 88.0 $\pm$ 7.0& \textbf{0.60 $\pm$ 0.22}& \textbf{1.61 $\pm$ 0.78}& 89.6 $\pm$ 5.8$^{bc}$& 0.53 $\pm$ 0.19$^{bc}$& 1.24 $\pm$ 0.52$^{bc}$\\

	    \underline{LVQuan19}&&&&&&&&&\\
	    
	    baseline & 85.7 $\pm$ 6.0& 0.75 $\pm$ 0.20& 1.91 $\pm$ 0.58& \textbf{87.4 $\pm$ 4.7}$^{coj}$& \textbf{0.67 $\pm$ 0.15}$^{oj}$& \textbf{1.76 $\pm$ 0.55}& 91.0 $\pm$ 4.2& 0.48 $\pm$ 0.18& 1.12 $\pm$ 0.51\\
		
		$L_{C_c}$&86.1 $\pm$ 5.6$^b$& \textbf{0.73 $\pm$ 0.19}$^{bo}$& \textbf{1.85 $\pm$ 0.56}$^{bo}$& 87.2 $\pm$ 4.7& \textbf{0.67 $\pm$ 0.15}$^{oj}$ & \textbf{1.75 $\pm$ 0.53}$^{oj}$& 91.5 $\pm$ 3.7$^b$& \textbf{0.45 $\pm$ 0.15}$^b$& \textbf{1.04 $\pm$ 0.40}$^{bj}$\\
		
		$L_{C_o}$&85.8 $\pm$ 6.6$^b$& 0.75 $\pm$ 0.24$^{b}$& 1.93 $\pm$ 0.78$^b$& 87.2 $\pm$ 4.7& 0.67 $\pm$ 0.16& 1.77 $\pm$ 0.56& \textbf{91.6 $\pm$ 5.3}$^{bcj}$& \textbf{0.46 $\pm$ 0.21}$^b$& \textbf{1.09 $\pm$ 0.70}$^{bj}$\\
		
        $L_{C_c}$+$L_{C_o}$&\textbf{86.2 $\pm$ 5.8}$^{bco}$& \textbf{0.74 $\pm$ 0.21}$^{bo}$& \textbf{1.86 $\pm$ 0.59}$^{bo}$& 87.3 $\pm$ 4.7$^c$& 0.67 $\pm$ 0.15& 1.76 $\pm$ 0.52& 91.5 $\pm$ 4.1$^b$& \textbf{0.45 $\pm$ 0.17}$^b$& 1.06 $\pm$ 0.46$^b$\\

	    \hline
	\end{tabular}
\end{table*}

Table \ref{tab:RegPost} reports the position error ($\Delta \textbf{c}$), orientation error ($\Delta \theta$) and the impact of shape prediction on landmark error ($\Delta \textbf{p}_b$) for the different consistency losses and different datasets. We observe that the landmark error due to shape coefficient errors is significantly decreased for all consistency losses and all datasets. For both $\Delta \textbf{c}$ and $\Delta \theta$, the three datasets show different results. Also here, the results are inconclusive about the optimal choice of consistency loss.

\begin{table*}[tb]
\centering
	\caption{Mean and standard deviation for position error ($\Delta \textbf{c}$), orientation error ($\Delta \theta$) and the impact of shape prediction on landmark error ($\Delta \textbf{p}_b$). Best values, for every dataset separately, are indicated in bold. Statistically significant improvement of every metric compared to baseline ($b$), $L_{C_c}$ ($c$), $L_{C_o}$ ($o$) or $L_{C_c}$+$L_{C_o}$ (joint, $j$) is indicated.
	}
	\scriptsize
	\label{tab:RegPost}
	\begin{tabular}{|l|l|l|l|l|l|l|l|l|l|}
		\hline
		 & \multicolumn{3}{c|}{IH} & \multicolumn{3}{c|}{ACDC} & \multicolumn{3}{c|}{LVQuan19}\\
		& \multicolumn{1}{c|}{$\Delta \textbf{c}$ [$pix$]} & \multicolumn{1}{c|}{$\Delta\theta$ [$^\circ$]} & \multicolumn{1}{c|}{$\Delta \textbf{p}_b$ [$pix$]}& \multicolumn{1}{c|}{$\Delta \textbf{c}$} & \multicolumn{1}{c|}{$\Delta\theta$ [$^\circ$]} & \multicolumn{1}{c|}{$\Delta \textbf{p}_b$ [$pix$]}& \multicolumn{1}{c|}{$\Delta \textbf{c}$ [$pix$]} & \multicolumn{1}{c|}{$\Delta\theta$ [$^\circ$]} & \multicolumn{1}{c|}{$\Delta \textbf{p}_b$ [$pix$]}\\
		\hline
		baseline& 0.82 $\pm$ 0.65& 5.29 $\pm$ 6.23& 0.66 $\pm$ 0.41&0.68 $\pm$ 0.50$^c$& \textbf{6.08 $\pm$ 6.17}& 0.56 $\pm$ 0.27&0.65 $\pm$ 0.41& 3.95 $\pm$ 3.31& 0.54 $\pm$ 0.17\\
		
		$L_{C_c}$& 0.81 $\pm$ 0.63& 5.29 $\pm$ 6.32& \textbf{0.64 $\pm$ 0.40}$^b$& 0.75 $\pm$ 0.51& \textbf{6.15 $\pm$ 6.30}& \textbf{0.55 $\pm$ 0.26}$^b$& 0.63 $\pm$ 0.40$^{bo}$& \textbf{3.79 $\pm$ 3.04}$^b$& \textbf{0.52 $\pm$ 0.17}$^{boj}$\\ % IP postproc

		$L_{C_o}$&\textbf{0.80 $\pm$ 0.65}$^b$& \textbf{5.18 $\pm$ 6.25}$^{bc}$& \textbf{0.64 $\pm$ 0.40}$^b$& \textbf{0.65 $\pm$ 0.48}$^{bcj}$& \textbf{6.07 $\pm$ 6.15}& \textbf{0.55 $\pm$ 0.28}$^b$& 0.67 $\pm$ 0.54& 3.82 $\pm$ 3.13& 0.54 $\pm$ 0.19$^{b}$\\ % TPS postproc

		$L_{C_c}$+$L_{C_o}$&\textbf{0.79 $\pm$ 0.63}$^{bc}$& \textbf{5.14 $\pm$ 6.05}$^{bc}$& \textbf{0.64 $\pm$ 0.39}$^b$&0.66 $\pm$ 0.47$^{c}$& \textbf{6.14 $\pm$ 6.27}& \textbf{0.55 $\pm$ 0.28}$^b$& \textbf{0.62 $\pm$ 0.41}$^{bco}$& \textbf{3.80 $\pm$ 3.06}$^{bo}$& 0.53 $\pm$ 0.18$^{b}$\\ % TPS postproc
		
	    \hline
	\end{tabular}
\end{table*}

In Table \ref{tab:physical}, MAE and correlation ($\rho$) of $A_{LV}$, $A_{MYO}$, $Dim_{LV}$ and $RWT$, obtained from the predicted shape coefficients with respect to the ground truth, are shown. In this table, $Dim_{LV}$ and $RWT$ were averaged over all segments. Additionally, the consistency between the metrics extracted from the 'Contour' or 'Map' representations is reported. Contrary to the results of 'Contour-Gt' in Table \ref{tab:RegSegPost} and $\Delta \textbf{p}_b$ in Table \ref{tab:RegPost}, we do not observe general improvement of predicted shape properties by using either consistency loss. For $L_{C_c}$, the highest number (12/12) of statistically significant best metrics was obtained, compared to 3/12 for the baseline. This was only 1/12 and 0/12 for $L_{C_o}$ and $L_{C_c}+L_{C_o}$, respectively. The consistency between 'Contour' and 'Map' representations improved for 29/36 metrics, showed no statistically significant difference for 5/36 metrics and deteriorated for 2/36 metrics. $L_{C_c}$ was best for 9/12 metrics, $L_{C_o}$ for 9/12 and $L_{C_c}+L_{C_o}$ for 7/12. Furthermore, $n_{err}$ was always lower when using a consistency loss compared to the baseline. 

\begin{figure*}[tb]
    \centering
    \includegraphics[width = \linewidth]{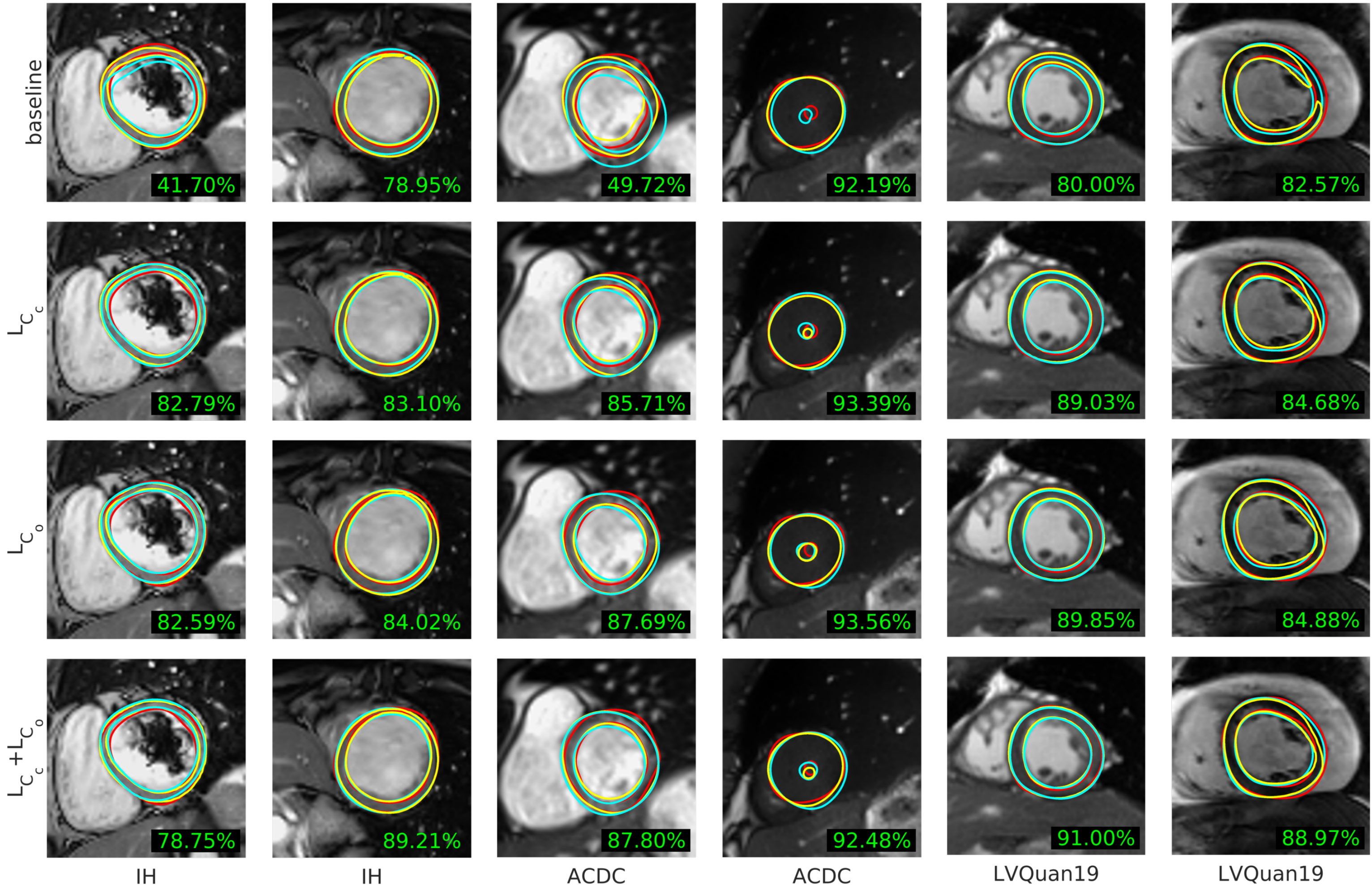}
    \caption{Example segmentations for the three datasets where the consistency loss was able to improve consistency between the segmentations obtained from the distance map ('Map', yellow) and the segmentations obtained from predicted shape and pose parameters ('Contour', cyan). The ground truth is depicted in red. The first example from every dataset shows and improved 'Contour' segmentation due to the consistency loss, while the second example shows an improved 'Map' segmentation. The DSC between 'Map' and 'Contour' is shown in green on the bottom right of every image. Cropped images are used for visualization purpose.}
    \label{fig:segresult_consistency}
\end{figure*}

Fig.~\ref{fig:segresult_consistency} shows some examples of improved segmentations due to the use of $L_{C_c}$ and/or $L_{C_o}$. The first example from every dataset shows improvement of the 'Contour' segmentation, while the second example shows an improvement of the 'Map' segmentation. The first example of the IH dataset shows a patient with an artificial valve. Interestingly, while no patients with artificial valves were included in the training set, the 'Map' segmentation shows good correspondence with the ground truth. The 'Contour' segmentation in contrast results in no overlap with the ground truth myocardium anteriorly and only slight overlap laterally, which is resolved after inclusion of $L_{C_c}$ and/or $L_{C_o}$. A similar error is seen in the example for dataset ACDC (infero-laterally) and to a smaller extent in the example for\linebreak dataset LVQuan19 (anteriorly). The second example of every dataset shows a case that was classified under $n_{err}$ in Table \ref{tab:physical} for the baseline experiment. For the examples of datasets IH and LVQuan19, the 'Map' segmentation results in an open myocardium, which is closed when using $L_{C_c}$ and/or $L_{C_o}$. However, the 'Contour' segmentation remains clearly superior for the example of dataset LVQuan19. In the example of dataset ACDC, both myocardium and LV cavity were segmented together for the baseline experiment and 'Map' output.  

\begin{table*}[tb]
    \centering
	\caption{MAE and correlation (MAE($\rho$)) for $A_{LV}$, $A_{MYO}$, $Dim_{LV}$ and $RWT$. Additionally, the number of cases with unrealistic shape in the 'Map' output ($n_{err}$) is reported. Best values, for every dataset separately, are indicated in bold. Statistically significant improvement of every metric compared to baseline ($b$), $L_{C_c}$ ($c$), $L_{C_o}$ ($o$) or $L_{C_c}$+$L_{C_o}$ (joint, $j$) is indicated.}
	\label{tab:physical}
	\scriptsize
\begin{tabular}{|l|l|l|l|l||l|l|l|l||c|}
		\hline
		& \multicolumn{4}{c||}{Contour-Gt} & \multicolumn{4}{c||}{Contour-Map} & Map\\
		& $A_{LV}$ [$mm^2$]& $A_{MYO}$ [$mm^2$]& $Dim_{LV}$ [$mm$]& $RWT$ [$mm$]& $A_{LV}$ [$mm^2$]& $A_{MYO}$ [$mm^2$]& $Dim_{LV}$ [$mm$] & $RWT$ [$mm$]& $n_{err}$\\
		\hline
		\underline{IH}& &&&&&&&&\\
	
		baseline& 110.96(0.99)& \textbf{117.65(0.96)}$^{oj}$& 1.88(0.99)& \textbf{1.10(0.88)}$^{oj}$ & 86.04(0.98)& 58.46(0.99)& 0.98(0.98)& 0.42(0.97)& 32\\
		
 		$L_{C_c}$& \textbf{99.68(0.99)}$^{boj}$& \textbf{118.46(0.96)}$^{oj}$& \textbf{1.77(0.99)}$^{boj}$& \textbf{1.11(0.88)}$^{oj}$& \textbf{74.41(0.98)}$^{bj}$& \textbf{48.69(0.99)}$^b$& 0.79(0.98)$^{b}$& \textbf{0.40(0.97)}$^{b}$ & 29\\
		
 		$L_{C_o}$&104.21(0.99)$^{bj}$& 123.22(0.96)& 1.80(0.99)$^b$& 1.14(0.88)&  \textbf{74.26(0.99)}$^{bj}$& \textbf{45.66(0.99)}$^b$& \textbf{0.75(0.99)}$^{bcj}$& \textbf{0.39(0.98)}$^{b}$ & 19 \\
		
 	    $L_{C_c}$+$L_{C_o}$& 107.12(0.99)& 121.19(0.96)& 1.82(0.99)$^b$& 1.13(0.88) & 80.93(0.98)& \textbf{45.62(0.99)}$^b$& 0.76(0.98)$^{bc}$& \textbf{0.39(0.98)}$^{b}$ &20 \\ % TPS postproc

 		\underline{ACDC}& && &&&&&&\\
	    
 		baseline& 102.04(0.99)& \textbf{133.25(0.97)}$^{oj}$& 1.87(0.99)& 1.27(0.91)$^{oj}$ & 81.99(0.99)& 69.99(0.98)& 1.08(0.98)& 0.62(0.97) & 10 \\
		
 		$L_{C_c}$& \textbf{95.09(0.99)}$^{boj}$& \textbf{130.58(0.97)}$^{oj}$& \textbf{1.81(0.99)}$^b$& \textbf{1.25(0.91)}$^{boj}$& \textbf{78.46(0.99)}$^{bj}$& \textbf{63.54(0.98)}$^b$& 1.03(0.98)$^b$& \textbf{0.59(0.98)}$^b$& 9\\ % IP postproc
		
 		$L_{C_o}$& 100.0(0.99)$^b$& 137.02(0.97)& \textbf{1.82(0.99)}$^b$& 1.29(0.91)& \textbf{79.32(0.99)}$^{bj}$& \textbf{63.04(0.99)}$^b$& \textbf{0.96(0.99)}$^{bc}$& \textbf{0.58(0.98)}$^b$ & 7 \\ % TPS postproc
		
 		$L_{C_c}$+$L_{C_o}$& 99.56(0.99)$^b$& 140.37(0.97)&\textbf{1.82(0.99)}$^b$& 1.29(0.92)& 83.31(1.00)& \textbf{63.05(0.99)}$^b$& \textbf{0.99(0.99)}$^{bc}$& \textbf{0.58(0.98)}$^b$ & 6 \\ % TPS 
 		
		\underline{LVQuan19}& && &&&&&&\\
		%gt op dezelfde manier berekend als hierboven
		baseline& 110.54(0.99)& 203.82(0.92)& 1.66(0.98)& 1.42(0.83)$^o$& 92.50(1.00)$^j$& 80.73(0.98)& 0.92(0.99)& 0.45(0.98)$^o$&5\\
		
		$L_{C_c}$& \textbf{106.37(0.99)}$^{boj}$& \textbf{198.07(0.92)}$^{boj}$& \textbf{1.57(0.99)}$^{boj}$& \textbf{1.38(0.84)}$^{boj}$& \textbf{88.56(1.00)}$^{boj}$& 72.93(0.99)$^b$& \textbf{0.85(1.00)}$^{bj}$& \textbf{0.43(0.98)}$^{bo}$&1\\ % IP postproc
		
		$L_{C_o}$& 108.67(0.99)& 207.65(0.91)& 1.62(0.98)$^{bj}$& 1.43(0.83)& 92.84(1.00)& 73.91(0.98)$^b$& \textbf{0.88(1.00)}$^{bj}$& 0.46(0.98)&3\\ % TPS postproc
		
		$L_{C_c}$+$L_{C_o}$& 111.53(0.99) & 199.66(0.93)$^{bo}$& 1.63(0.99)& 1.40(0.80)$^{o}$& 99.65(1.00)$^j$& \textbf{67.12(0.99)}$^{bco}$& 0.91(1.00)& \textbf{0.44(0.98)}$^{o}$&1 \\ % TPS postproc
		
	    \hline

\end{tabular}
\end{table*}

Tables~\ref{tab:litCompGeo}~and~\ref{tab:litCompPhysical} show the results of the comparison of our method with the state-of-the-art. Table \ref{tab:litCompGeo} reports DSC, MBE and HD for the network with skip connections and for the four nnUNet variations. While adding skip connections significantly improved the 'Map' output, it significantly deteriorated the\linebreak 'Contour' output. For nnUNet, we observed that our preprocessing, including image resampling and intensity cropping, significantly deteriorated the results. The statistically significantly best DSC and MBE were obtained when ensembling the predictions of the 2D and 3D nnUNet CNNs which is the approach proposed by \cite{Isensee} and \cite{Isensee2021}. In contrast, the statistically significantly best HD was achieved by our baseline method with skip connections in the 'Map' output.

\begin{table}[tb]
\centering
	\caption{Effect of skip connections (SK) on our baseline method and comparison with state-of-the-art segmentation method nnUNet (\cite{Isensee2021}) for dataset ACDC. For nnUNet, results obtained on images identically preprocessed as in our method (2D preproc) and results obtained from the raw dataset using a 2D or 3D CNN and the combination of both predictions (2D+3D) are given. Mean and standard deviation for DSC, MBE and HD are reported. Best values are indicated in bold. Statistically significant improvement of every metric compared to baseline ($b$), $L_{C_c}$ ($c$), $L_{C_o}$ ($o$), $L_{C_c}$+$L_{C_o}$ (joint, $j$), our baseline method with skip connections ('Map' output, $s$), 2D nnUNet with preprocessing ($p$), 2D nnUNet ($2$), 3D nnUNet ($3$) or 2D+3D nnUNet (ensemble, $e$) is indicated. Additionally, the 'Map-Gt' results that showed a statistically significantly improvement compared to the 'Contour-Gt' results are indicated with gray background. Statistically significant deterioration of the 'Contour-Gt' output when adding skip connections is indicated with '$-$'.}
	\scriptsize
	\label{tab:litCompGeo}
	\begin{tabular}{|l|l|l|l|}
		\hline
		ACDC &  DSC [$\%$] & MBE [pix]& HD [pix] \\
		\hline
		\multicolumn{4}{|l|}{baseline SK}\\
		\hline
		Contour-Gt & 83.2 $\pm$ 9.4$^-$& 0.80 $\pm$ 0.28$^-$& 2.13 $\pm$ 0.86$^-$\\
		Map-Gt &\cellcolor{gray!20}89.9 $\pm$ 5.5$^{bcoj}$& \cellcolor{gray!20}0.53 $\pm$ 0.17$^{bcoj}$& \cellcolor{gray!20}\textbf{1.48 $\pm$ 0.73}$^{bcojp23e}$\\
		\hline
		\multicolumn{4}{|l|}{nnUNet (\cite{Isensee2021})}\\
		\hline
		2D preproc &89.8 $\pm$ 8.1$^{bcojs}$& 0.44 $\pm$ 0.23$^{bcojs}$& 1.71 $\pm$ 1.96\\ % HD is statistically significantly the same compared to bcoj; HD is statistically significantly worse compared to s
		2D &90.0 $\pm$ 8.1$^{bcojsp}$& 0.42 $\pm$ 0.25$^{bcojsp}$& 1.63 $\pm$ 1.43$^{cojp}$\\ % no difference with HD baseline, HD significantly worse compared to s
		3D & 90.2 $\pm$ 5.8$^{bcojsp}$& 0.42 $\pm$ 0.18$^{bcojsp}$& 1.70 $\pm$ 1.46$^p$\\ % no difference in HD,HD significantly worse compared to s
		2D+3D &\textbf{90.6 $\pm$ 6.9}$^{bcojsp23}$& \textbf{0.40 $\pm$ 0.21}$^{bcojsp23}$& 1.55 $\pm$ 1.03$^{bcojp23}$\\ % HD significantly worse compared to s
		
	    \hline
	\end{tabular}
\end{table}

\begin{figure}[tb]
    \centering
    \includegraphics[width = 0.8\linewidth]{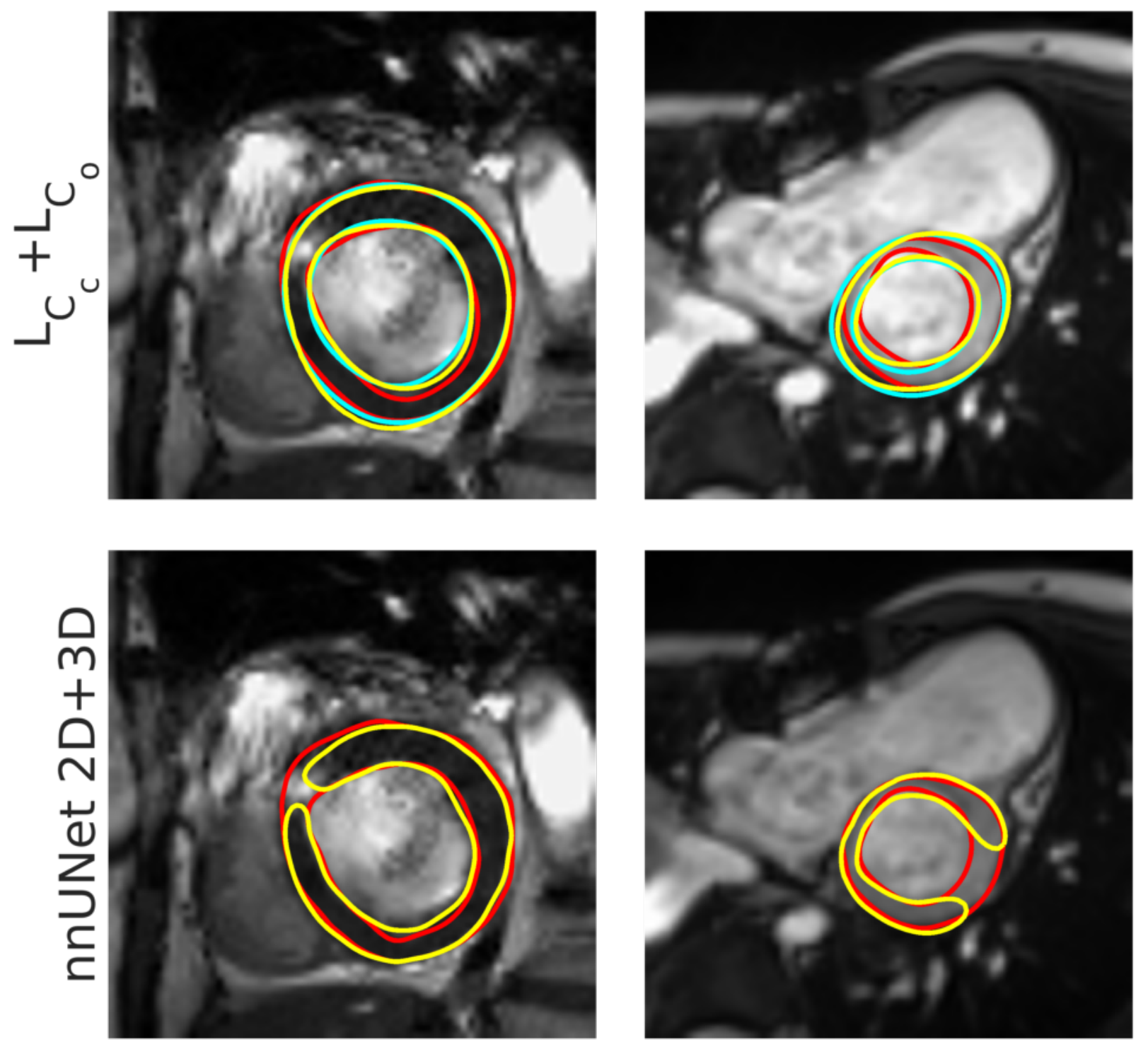}
    \caption{nnUNet segmentations with discontinuous myocardium and the corresponding results with our proposed method $L_{C_c}$+$L_{C_o}$. Red contours show the ground truth. For nnUNet, predictions are indicated in yellow. For $L_{C_c}$+$L_{C_o}$, both results obtained from distance map prediction ('Map', yellow) and from the predicted shape and pose parameters ('Contour', cyan) are shown.}
    \label{fig:segresult_nnunet}
\end{figure}

Table \ref{tab:litCompPhysical} reports MAE and correlation for the four LV parameters $A_{LV}$, $A_{MYO}$, $Dim_{LV}$ and $RWT$ using more standard approaches compared to our shape parameter prediction. For all three datasets, the errors on the metrics derived from the 'Map' output and the predicted orientation show statistically significantly superior or equal results compared to the 'Contour' output. In accordance with Table \ref{tab:litCompGeo}, the addition of skip connections deteriorated the 'Contour' output and improved the 'Map' output for dataset ACDC. Furthermore, $Dim_{LV}$ and $RWT$ were significantly better for the baseline network with skip connections compared to the 2D nnUNet with identical preprocessing, while $A_{LV}$ and $A_{MYO}$ did not show a difference. The statistically significantly best results for the ACDC dataset were obtained using the 2D+3D nnUNet for all four parameters. However, we observe a higher number of myocardial predictions with unrealistic shape ($n_{err}$) compared to our proposed method. Two examples are given in Fig.~\ref{fig:segresult_nnunet}. Additionally, $Dim_{LV}$ and $RWT$ were not calculated for 48 slices (=3$\%$ of the data) for the proposed 2D+3D nnUNet method due to difficulties to extract the orientation.

Table \ref{tab:litCompPhysical} also shows the metrics of the top three entries of the LVQuan19 challenge (\url{https://lvquan19.github.io/}) as reported in \cite{Acero}, \cite{Gessert2020} and \cite{Tilborghs2020a}. Our method ($L_{C_c}$, $L_{C_o}$ and $L_{C_c}$+$L_{C_o}$ 'Contour' output) ranked 2/4 for  $Dim_{LV}$ and 3/4 for $A_{LV}$, $A_{MYO}$ and $RWT$ compared to these methods.   

\begin{table*}[tb]
    \centering
	\caption{Comparison with other methods to calculate $A_{LV}$, $A_{MYO}$, $Dim_{LV}$ and $RWT$. For the IH dataset, the metrics calculated from the 'Map' output are reported. For the ACDC dataset, the metrics calculated from the 'Map' output, from our baseline method where skip connections (SK) had been added and from the multi-class state-of-the-art segmentation nnUNet (\cite{Isensee2021}), are given. For the LVQuan19 dataset, the metrics calculated from the 'Map' output as well as the results from the top three entries of the challenge (\url{https://lvquan19.github.io/}) are given. All metrics are reported as MAE and correlation (MAE($\rho$)). Additionally, the number of cases with unrealistic myocardial shape in the 'Map' output ($n_{err}$) and the number of cases where the cardiac orientation could not be calculated ($n_{err,\theta}$) are reported. Best values, for every dataset separately, are indicated in bold. Statistically significant improvement of every metric compared to baseline ($b$), $L_{C_c}$ ($c$), $L_{C_o}$ ($o$), $L_{C_c}$+$L_{C_o}$ (joint, $j$), our baseline method with skip connections ('Map' output, $s$), 2D nnUNet with preprocessing ($p$), 2D nnUNet ($2$), 3D nnUNet ($3$) or 2D+3D nnUNet (ensemble, $e$) is indicated. Additionally, the 'Map-Gt' results that showed a statistically significantly improvement compared to the 'Contour-Gt' results are indicated with gray background. Statistically significant deterioration of the 'Contour-Gt' output when adding skip connections is indicated with '$-$'. $^1$The results shown for \cite{Acero} included only 12 subjects instead of 56, $^2$in \cite{Gessert2020}, the average MAE of LV and myocardial area was reported to be 111$mm^2$ with a correlation of 0.98, $^3$place 3 was obtained by \cite{Tilborghs2020a}.} 
	\label{tab:litCompPhysical}
	\scriptsize
\begin{tabular}{|l|l|l|l|l|l|l||c|c|}
		\hline
		&&& $A_{LV}$ [$mm^2$]& $A_{MYO}$ [$mm^2$]& $Dim_{LV}$ [$mm$]& $RWT$ [$mm$] & $n_{err}$ & $n_{err,\theta}$\\
		\hline
		\hline
		IH & Map-Gt & baseline &\cellcolor{gray!20}88.71(0.97)& \textbf{105.53(0.94)}& \cellcolor{gray!20}\textbf{1.59(0.97)}& \cellcolor{gray!20}\textbf{1.03(0.89)}$^o$&32&\\
		
		&&$L_{C_c}$& 90.57(0.97)& \textbf{106.30(0.94)}& \cellcolor{gray!20}\textbf{1.59(0.97)}$^o$& \cellcolor{gray!20}\textbf{1.03(0.89)}$^o$&29&\\
		
		&&$L_{C_o}$& \textbf{89.83(0.98)}$^{bc}$& \cellcolor{gray!20}\textbf{107.43(0.95)}& \cellcolor{gray!20}1.61(0.98)& \cellcolor{gray!20}1.05(0.89)&19&\\
		
		&&$L_{C_c}$+$L_{C_o}$& \cellcolor{gray!20}\textbf{88.96(0.98)}$^{bc}$& \cellcolor{gray!20}\textbf{106.92(0.95)}& \cellcolor{gray!20}\textbf{1.60(0.98)}$^o$& \cellcolor{gray!20}\textbf{1.03(0.89)}$^o$&20&\\
		
		\hline
		\hline
		ACDC & Map-Gt & baseline & \cellcolor{gray!20}79.60(0.99)& 130.30(0.96)& \cellcolor{gray!20}1.52(0.98)& \cellcolor{gray!20}1.18(0.92)& 10&\\
		
		&&$L_{C_c}$&  \cellcolor{gray!20}81.01(0.99)& \cellcolor{gray!20}125.03(0.96)$^b$& \cellcolor{gray!20}1.50(0.98)$^b$& \cellcolor{gray!20}1.16(0.92)$^b$ & 9&\\
		
		&&$L_{C_o}$& \cellcolor{gray!20}75.63(0.99)$^{bcj}$& \cellcolor{gray!20}125.72(0.97)$^b$& \cellcolor{gray!20}1.50(0.98)$^b$& \cellcolor{gray!20}1.17(0.92)$^b$& 7&\\
		
		&&$L_{C_c}$+$L_{C_o}$& \cellcolor{gray!20}80.36(0.99)& \cellcolor{gray!20}126.08(0.97)$^b$& \cellcolor{gray!20}1.51(0.98)$^b$& \cellcolor{gray!20}1.16(0.92)$^b$& 6&\\
		
		\cline{3-9}
 		& baseline SK & Contour-Gt& 112.68(0.99)$^-$& 146.62(0.96)$^-$& 2.14(0.98)$^-$& 1.41(0.89)$^-$ &&\\ 
		&&Map-Gt & \cellcolor{gray!20}64.10(0.99)$^{bcoj}$& \cellcolor{gray!20}111.87(0.97)$^{bcoj}$& \cellcolor{gray!20}1.24(0.99)$^{bcojp}$& \cellcolor{gray!20}1.00(0.94)$^{bcojp}$ & 8&\\
		\cline{3-9}
		&nnUNet&2D preproc & 67.45(1.00)$^{bcoj}$& 118.52(0.98)$^{bcoj}$& 1.19(0.96)$^{bcoj}$& 1.00(0.92)$^{bcoj}$ &14 &71 \\ % voor lv area and myo area no difference with SK, SK dim and rwt beter dan nnunet; verschil in stat sign vooral te verklaren door van welke hoeveelheid beelden metric berekend werd: areas werden voor 14 beelden niet berekend, dim en rwt voor 71!
		
		 &(\cite{Isensee2021})&2D &60.56(1.00)$^{bcojsp}$& 98.53(0.98)$^{bcojsp}$& 1.11(0.96)$^{bcojp}$& 0.92(0.93)$^{bcojsp}$&14 &53\\ % dim no statistical difference with s
 		&&3D & 61.18(1.00)$^{bcojsp}$& 96.93(0.98)$^{bcojsp}$& 1.15(0.97)$^{bcojsp}$& 0.90(0.93)$^{bcojsp2}$&13 & 45\\
 		&&2D+3D& \textbf{57.98(1.00)}$^{bcojsp23}$& \textbf{92.11(0.98)$^{bcojsp23}$}& \textbf{1.07(0.97)}$^{bcojsp23}$& \textbf{0.87(0.94)}$^{bcojsp23}$&13 &48\\
 		
 		\hline
 		\hline
		LVQuan19&Map-Gt&baseline& 113.70(0.99)$^{coj}$& \cellcolor{gray!20}184.88(0.93)$^{oj}$& \cellcolor{gray!20}\textbf{1.51(0.99)}$^{coj}$& \cellcolor{gray!20}1.35(0.85)$^{oj}$&5& \\
		
		&&$L_{C_c}$& 120.20(0.99)& 183.93(0.93)$^{oj}$& 1.54(0.99)& \cellcolor{gray!20}1.35(0.85)$^{oj}$&1&\\
		
		&&$L_{C_o}$& 118.17(0.99)$^{c}$& \cellcolor{gray!20}186.84(0.93)& 1.55(0.99)& \cellcolor{gray!20}1.36(0.84)&3&\\
		
		&&$L_{C_c}$+$L_{C_o}$& 117.28(0.99)$^{c}$& 186.91(0.93)& \cellcolor{gray!20}1.53(0.99)& \cellcolor{gray!20}1.36(0.85)&1&\\
		
		\cline{3-9}
		&Challenge &Place 1$^{1}$ &\textbf{92(-)}&\textbf{121(-)}&1.52(-) &\textbf{1.01(-)}&&\\
 		&(\url{https://lvquan19.github.io/})&Place 2$^2$ & '111(0.98)'&'111(0.98)'& 1.84(0.98)&1.22(0.86)&&\\
 		&&Place 3$^3$ & 186(0.97)&222(0.88)&3.03(0.95) &1.67(0.73)&&\\
	    \hline

\end{tabular}
\end{table*}

\section{Discussion}
In this paper, we presented a CNN for direct prediction of shape and pose parameters of the myocardium in cardiac SA MR images. Compared to conventional semantic segmentation where segmentation is formulated as a per-pixel classification problem, the predicted shape coefficients are directly linked to a landmark-based representation and as such allow straightforward calculation of regional shape properties. Furthermore, the shape model was learned from the training set itself and represent the main modes of shape variation in the dataset, helping the CNN in predicting realistic myocardial shapes. As stated in the introduction, similar architectures have been used in literature to perform joint regression and segmentation. However, compared to the previous methods applying this joint strategy, the predicted parameters in our approach belong to a statistical shape model. One of the advantages of our method is that this underlying model yields a corresponding segmentation and hence provides direct visual feedback to assess the quality of parameter regression. Furthermore, regression of a comprehensive shape model offers flexibility in the LV parameters that can be quantified as different parameters can be calculated from the predicted shape without need for retraining of the CNN. Additionally, our approach assures consistency between the different predicted LV parameters, which is not the case for methods where the different parameters, for example myocardial area and $RWT$, are predicted directly with a CNN.

In our experiments (Table \ref{tab:Seg}), we found that the regression of shape and pose parameters is a more difficult task compared to semantic segmentation. Indeed, when performing semantic segmentation a prediction is made for every pixel, resulting in training on and prediction of a large number of correlated values per image (128$\times$128 = 16,384 in our case). In contrast, shape and pose regression required the estimation of a limited number (15) of uncorrelated values from relatively few training data. The training data was artificially increased by using model-guided shape augmentation. This approach proved beneficial for the prediction of shape coefficients, LV center prediction and orientation but did not improve semantic segmentation compared to conventional augmentation of position and orientation. We also observed in Fig.~\ref{fig:shape} that higher coefficients are more difficult to predict accurately, which is not surprising since these coefficients represent only small shape variations (Fig.~\ref{fig:shapemodel}).

To assist the training of the CNN weights, we appended an additional pathway for semantic segmentation. Consequently, the layers used jointly for parameter regression and semantic segmentation receive information from both tasks during training. The results in Table \ref{tab:Seg} showed that the addition of this pathway successfully improved the segmentation obtained from predicted shape and pose parameters. We hypothesize that this superior result is achieved because the semantic segmentation task is able to learn more meaningful geometric features in the common layers. To enforce consistency between the semantic segmentation output ('Map') and the segmentation obtained from predicted shape and pose parameters ('Contour'), we introduced two additional loss functions: one distance-based loss and one overlap-based loss. While both losses were separately able to significantly improve the consistency between 'Map' and 'Contour' predictions (Tables~\ref{tab:RegSegPost}~and~\ref{tab:physical}) and DSC, MBE, HD and $\Delta\textbf{p}_b$ obtained from the 'Contour' output (Tables~\ref{tab:RegSegPost}~and~\ref{tab:RegPost}), combining the two did not further improve the results. We hypothesize that the two losses are not able to reinforce each other since their combination results in local minima due to their different evolution during training. The consistency between 'Map' and 'Contour' segmentations (Table \ref{tab:RegSegPost}) showed a preference for $L_{C_o}$, while the results on $A_{LV}$, $A_{MYO}$, $Dim_{LV}$ and $RWT$ with respect to the ground truth (Table \ref{tab:physical}) were best for $L_{C_c}$, since with $L_{Co}$, we observed a slight oversegmentation of the myocardium.   

Instead of conventional semantic segmentation where a per-pixel probability is predicted to belong to a certain class, we formulated the semantic segmentation problem as the regression of signed distance maps for two reasons: (1) recent literature has shown that the use of distance maps is beneficial for semantic segmentation (\cite{
Karimi2020,Kervadec2018,Xue2019,Dangi2019,Navarro,Ma2020}) and (2) prediction of signed distance maps allows straightforward implementation of our $L_{C_c}$ consistency loss. Inclusion of distance-based loss functions by using distance maps, in combination with an overlap-based metric, has been implemented in many forms, as discussed in the introduction. In our work, $L_{D}$ was defined as the sum of DSC loss and mean squared error between true and predicted distance map. Contrary to the losses defined by e.g. \cite{Karimi2020} and \cite{Kervadec2018}, our loss was not derived from commonly used metrics to validate segmentation performance such as MBE and HD. Instead, it trains the network to predict the signed distance maps as accurately as possible for the complete image. Nonetheless, from the results in Table \ref{tab:Seg} a similar conclusion is drawn as for other distance-based metrics (\cite{Ma2020,Karimi2020,Li2020}): the use of distance maps causes a significant decrease in HD, as well as decreasing the number of unrealistic shapes, while DSC is mainly unaffected. 

We applied our method to two public datasets, the ACDC challenge data (\cite{Bernard2018}) and the LVQuan19 challenge data (\url{https://lvquan19.github.io/}). For the ACDC dataset, we also performed a multi-class segmentation using the public nnUNet code (\url{https://github.com/MIC-DKFZ/nnUNet}) which has won both the ACDC challenge (\cite{Bernard2018,Isensee}) and the Multi-Centre, Multi-Vendor $\&$ Multi-Disease Cardiac Image Segmentation Challenge (M$\&$Ms, \cite{Full2021}). This method automatically defines appropriate preprocessing and training parameters based on data characteristics using look-up tables. The results in Tables~\ref{tab:litCompGeo} and \ref{tab:litCompPhysical} indicate better performance compared to our approach. However, HD is superior for the 'Map'-output of our CNN (Table \ref{tab:litCompGeo}). We believe that this is mainly due to the higher number of erroneous shapes for nnUNet (see Table~\ref{tab:litCompPhysical} and Fig.~\ref{fig:segresult_nnunet}). The decrease in $n_{err}$ with our CNN is due to the use of distance maps on the one hand (Table~\ref{tab:Seg}) and the use of consistency losses $L_{C_c}$ and $L_{C_o}$ on the other hand (Table~\ref{tab:physical}).

Two key differences between nnUNet and the ‘Map’-output of our CNN are the use of skip connections and the used preprocessing. Although skip connections improved the ‘Map’-output, it simultaneously also decreased the performance for the ‘Contour’-output (see Tables~\ref{tab:litCompGeo} and \ref{tab:litCompPhysical}). Furthermore, we resampled the input images to a pixel size of 2$mm$x2$mm$, which is larger than the 1.25$mm$x1.25$mm$ used by \cite{Isensee}, since we noticed that the prediction of shape and pose parameters was largely influenced by batch size, where a large batch size, permitted by the use of a larger pixel size, resulted in better predictions (results not shown). The impact of batch size was more pronounced than the impact of an increase in pixel size from 1$mm$x1$mm$ to 2$mm$x2$mm$. However, we believe that the shape and pose prediction can be further improved when more GPU memory would be available, or when a two-step approach consisting of a detection CNN and a shape and pose prediction CNN using cropped high-resolution images would be used.

In this paper, we proposed to calculate the LV parameters ($A_{LV}$, $A_{MYO}$, $Dim_{LV}$ and $RWT$) using the shape model and predicted shape parameters. Consequently, the regional LV parameters $Dim_{LV}$ and $RWT$ could be obtained with a simple subtraction of two landmarks. However, better results were achieved by extracting these parameters from a predicted segmentation map ('Map'-output or nnUNet), which is in accordance with the fact that the winner of the LVQuan19 challenge (\cite{Acero}) was the only segmentation-based method in this challenge. In contrast to the contour-based method, a segmentation-based method requires a (smooth) contour extraction and an estimation of cardiac orientation for $RWT$ and $Dim_{LV}$, which can be obtained either by predicting the cardiac orientation as an additional output in a multi-task CNN ('Map'-output and \cite{Vigneault2018}) or by performing multi-class semantic segmentation (nnUNet and \cite{Isensee,Ruijsink2020}). Furthermore, such method might occasionally result in erroneous segmentations for which it is difficult to calculate the LV parameters ($n_{err}$). We also want to stress the high number of slices for which the regional metrics, $Dim_{LV}$ and $RWT$, could not be calculated from the nnUNet predictions with the same method as we used to extract the LV parameter ground truths ($n_{err,\theta}$). Such failure cases are avoided by our contour-based method. Finally, we want to highlight again that parameter prediction is much more sensitive to the dataset size compared to semantic segmentation which might be an explanation for the superior results of the segmentation branch.

Another option to calculate regional shape parameters from a statistical shape model would be to post-process the predicted segmentation with a contour fitting step. However, in this paper we intended to create an end-to-end trainable model which can be considered to be more elegant. Furthermore, our approach allows to omit the segmentation pathway during prediction which makes this end-to-end approach more efficient.

Our loss function (Eq. \ref{eq:totalloss}) contains nine parameters ($\gamma_{b}$, $\gamma_{\Phi}$, $\gamma_{p}$, $\gamma_{D}$, $\gamma_{C_c}$, $\gamma_{C_o}$, $\mu_{\Phi}$, $\mu_{D}$ and $\alpha$) which were determined experimentally on the first fold of dataset IH. As can be noted in Fig.~\ref{fig:consistencyloss_ip_analysis}~and~\ref{fig:consistencyloss_warp_analysis} (Supplementary Material) for $\gamma_{C_c}$ and $\gamma_{C_o}$, respectively, the performance of the CNN is highly dependent on an appropriate choice for each parameter in the loss function. The optimal weights might not have been attained through our time-consuming experimental tuning. In this respect, the training might benefit from a more principled approach for weight tuning, such as the method of \cite{Kendall} where uncertainty is used to weight multiple losses.   

In our method, $L_{C_o}$ was obtained using a thin-plate-spline deformation field calculated from $\overline{\textbf{s}}$ and $\textbf{p}_p$ to transform $D_p$ to the image space of the average shape. This approach was chosen since the calculation of exact distance maps is computationally intensive and non-differentiable (\cite{Karimi2020}). However, other options to generate $S^{b,\Phi}_p$ itself exist. In this respect, \cite{Wang2020} propose a CNN for distance map generation from shape coefficients. A disadvantage is that a method using a trained model risks to propagate model imperfections. \cite{Leveton2000} create a statistical model from signed distance maps instead of from a landmark based representation. Such model would allow to create a distance map from shape coefficients with a simple linear function, although the linear combination of distance maps does not result in a true distance map.

We trained and validated a single network to handle multiple slice positions (apex to base) and multiple time points (ED and ES) on datasets IH and ACDC, and on all time points of a mid-cavity slice on dataset LVQuan19. This implies that the statistical shape model not only captures subject variability but also time variation and variation of myocardial shape appearing in different 2D SA slices from apex to base. The advantage of modeling all these variations simultaneously is a richer dataset for training the CNN. The cases used for training and validating our method contained closed, separated endo- and epicardial contours. However, apical and basal slices might contain deviating shapes. In our method, an apical segmentation that only contains a single epicardial contour could, for example, be represented by 18 landmarks on the epicardium and 18 coinciding landmarks at the center. For a basal slice where the myocardium is not a full circle, two coinciding contours can indicate the boundary of LV cavity not surrounded by myocardium. While our method can thus be extended with these such alternative shapes, their prediction might suffer from their lesser occurrence compared to mid-cavity shapes. Therefore, we believe that it is more valuable to direct further research towards a 3D version of our approach. Since the parameter regression showed to be more sensitive to the number of samples, we expect a larger number of datasets to be required.

Our method is trained to segment the myocardium, but we also extracted parameters of the LV cavity ($A_{LV}$ and $Dim_{LV}$) for which we did not optimize a loss during training. A loss on the LV cavity could be included by extending our approach to a multi-class CNN predicting multiple distance maps, including background, simultaneously. After prediction, a single class for every pixel could be determined by adding a softmax activation function after the soft binarization operation of Eq.~\ref{eq:sigmoidconversion}. In accordance with the segmentation loss $L_D$ (Eq.~\ref{eq:segloss}) used in this paper, the segmentation loss would be a combination of MSE loss on the distance maps and multi-class Dice loss on the binarized segmentation maps. Such approach could also be applied to a multi-class segmentation problem that includes LV, myocardium and RV in a single statistical model.

This paper described a shape constrained approach for myocardium segmentation using a CNN in 2D SA images but this concept can as well be applied to other segmentation tasks. Examples are joint modeling of myocardium, LV and RV, 3D cardiac segmentation or application of the method on other clinical data, e.g. hippocampus segmentation in brain images. 

\section{Conclusion}
In this paper, we presented a CNN for direct prediction of shape and pose parameters of the myocardium in cardiac SA MR images. Compared to conventional semantic segmentation, the predicted shape coefficients are directly linked to an oriented landmark-based representation and as such allow straightforward calculation of regional shape properties. Experiments showed an increase in performance when semantic segmentation was used to guide robust parameter prediction. Enforcing explicit consistency between parameter and segmentation outputs by newly defined loss functions is promising and no clear preference for a distance-based loss function or an overlap-based loss function was found. 

\section*{Acknowledgment}
Sofie Tilborghs is supported by a Ph.D. fellowship of the Research Foundation - Flanders (FWO). This work is partially funded by KU Leuven Internal Funds C24/18/047 (promotor F. Maes) and also received funding from the Flemish Government under the “Onderzoeksprogramma Artifici\"{e}le Intelligentie (AI) Vlaanderen” programme.

%%Harvard
\bibliographystyle{model2-names.bst}\biboptions{authoryear}
\bibliography{FullPaperCNNestimateShapeModel_noURL,Bibliography}

\newpage
\section*{Supplementary Material}

\begin{figure}[b]
    \centering
    \includegraphics[width = \linewidth]{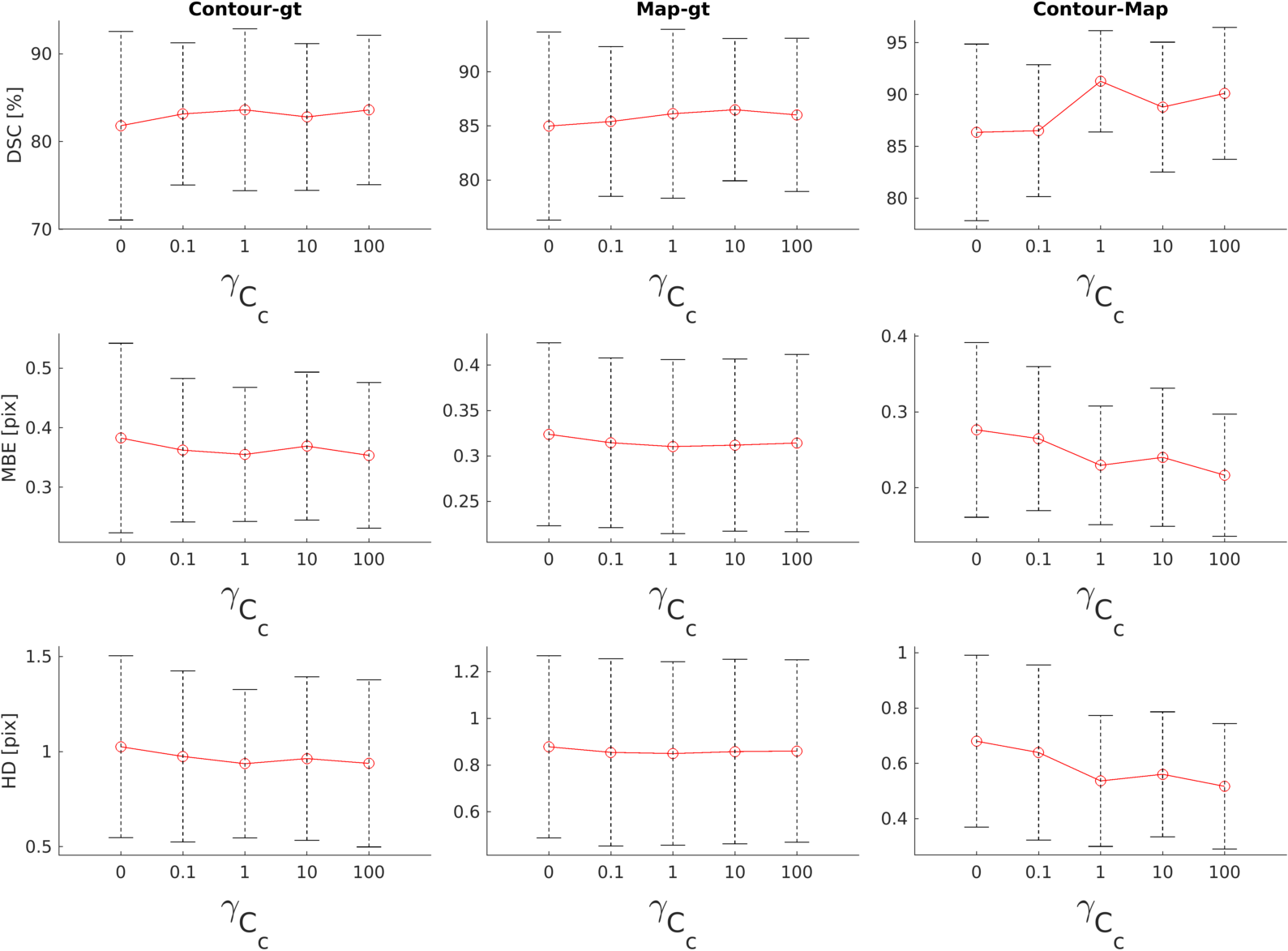}
     \caption{Evolution of DSC, MBE and HD in function of $\gamma_{C_c}$ for segmentations reconstructed using shape and pose predictions ('Contour') or segmentations obtained from distance map prediction ('Map'). The first two columns ('Contour-gt' and 'Map-gt') show the results with respect to ground truth segmentations. The third column compares the two predicted representations ('Contour-Map').}
    \label{fig:consistencyloss_ip_analysis}
\end{figure}

\begin{figure}[b]
    \centering
    \includegraphics[width = \linewidth]{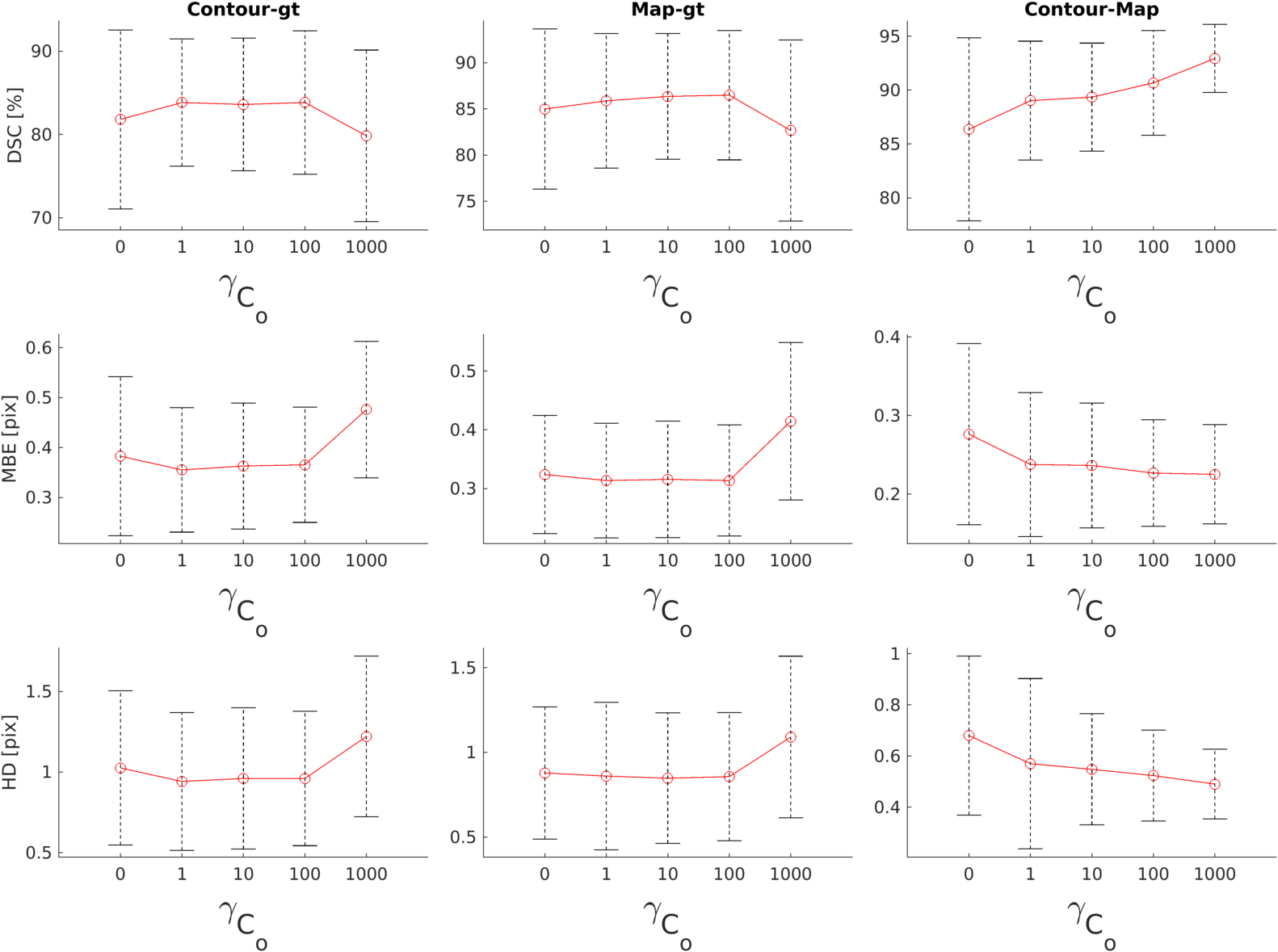}
    \caption{Evolution of DSC, MBE and HD in function of $\gamma_{C_o}$ for segmentations reconstructed using shape and pose predictions ('Contour') or segmentations obtained from distance map prediction ('Map'). The first two columns ('Contour-gt' and 'Map-gt') show the results with respect to ground truth segmentations. The third column compares the two predicted representations ('Contour-Map').}
    \label{fig:consistencyloss_warp_analysis}
\end{figure}

Fig. \ref{fig:consistencyloss_ip_analysis} and Fig. \ref{fig:consistencyloss_warp_analysis} show the effect of different values of $\gamma_{C_c}$ and $\gamma_{C_o}$ on the segmentations. As intended, $L_{C_c}$ minimizes the boundary error between 'Contour' and 'Map' predictions, which can be observed as a decrease in MBE and HD for increasing $\gamma_{C_c}$ in the last column of Fig. \ref{fig:consistencyloss_ip_analysis}. Furthermore, we see a slight improvement for DSC of both 'Contour-Gt' and 'Map-Gt' and for MBE of 'Contour-Gt' for all $\gamma_{C_c}>0$. The last column of Fig. \ref{fig:consistencyloss_warp_analysis} shows an increase in DSC and decrease in MBE and HD between 'Contour' and 'Map' outputs for increasing $\gamma_{C_o}$, representing a higher consistency. DSC, MBE and HD calculated from the 'Contour' outputs improve by adding the loss $L_{C_o}$ with small $\gamma_{C_o}$ ($\gamma_{C_o}=1$) and remain relatively stable for $\gamma_{C_o}=10$ and $\gamma_{C_o}=100$ before strongly degrading for $\gamma_{C_o}=1000$. In conclusion, $L_{C_c}$ and $L_{C_o}$ can both improve the consistency between 'Contour' and 'Map' and improve the segmentation performance of 'Contour' while the performance of 'Map' remains relatively unaffected. However, exact consistency can not be attained because higher consistency will result in worse segmentation results for both 'Contour' and 'Map' outputs. Based on these results, $\gamma_{C_c}=1$ and $\gamma_{C_o}=10$ were chosen to train the 5-fold cross validation on the three datasets.

%Supplementary material that may be helpful in the review process should
%be prepared and provided as a separate electronic file. That file can
%then be transformed into PDF format and submitted along with the
%manuscript and graphic files to the appropriate editorial office.

\end{document}